\documentclass[reqno, 11pt, a4paper]{amsart}
\usepackage[dvipsnames]{xcolor}
\usepackage{tikz-cd}
\usepackage[utf8]{inputenc}
\usepackage[T1]{fontenc}
\usepackage{amsmath}
\usepackage{amsfonts,amssymb}
\usetikzlibrary{matrix}
\usepackage{tensor}
\usepackage{bbm}
\usepackage{graphicx}
\graphicspath{{./pics_BTM/}{./BTM_graphs/}}

    \newcommand{\viq}{\raisebox{-.32\height}{\includegraphics[height=2.8ex]{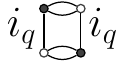}}}

\newcommand{\AppGr}[1]{\hspace{-1ex}\includegraphics[height=2.7cm]{BTM_graphs/QuadraticPillows_model_Grx#1}\hspace{-1ex}}
\newcommand{\maketrace}[1]{\Tr_{#1}(T,\bar T)}
    \newcommand{\vione}{\raisebox{-.32\height}{\includegraphics[height=2.8ex]{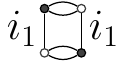}}}

\newcommand{\Y}{ \raisebox{-.38\height}{  \includegraphics[height=3.39ex]{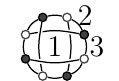}}}

\newcommand{\tenQ}{ \raisebox{-.318\height}{  \includegraphics[height=3.399ex]{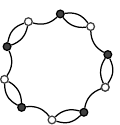}}}
\newcommand{\tens}{ \raisebox{-.318\height}{  \includegraphics[height=3.399ex]{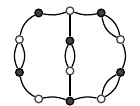}}}
\newcommand{\tent}{ \raisebox{-.318\height}{  \includegraphics[height=3.399ex]{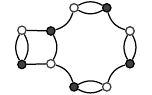}}}
\newcommand{\tend}{ \raisebox{-.318\height}{  \includegraphics[height=3.399ex]{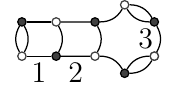}}}
\newcommand{\teny}{ \raisebox{-.318\height}{  \includegraphics[height=3.399ex]{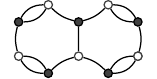}}}
\newcommand{\tene}{ \raisebox{-.318\height}{  \includegraphics[height=3.399ex]{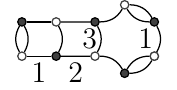}}}

\newcommand{\X}{ \raisebox{-.348\height}{  \includegraphics[height=2.99ex]{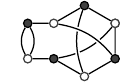}}}

\newcommand{\Xu}{ \raisebox{-.48\height}{  \includegraphics[height=2.99ex]{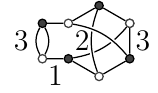}}}
\newcommand{\Xd}{ \raisebox{-.48\height}{  \includegraphics[height=2.99ex]{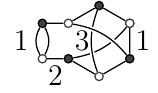}}}
\newcommand{\Xt}{ \raisebox{-.48\height}{  \includegraphics[height=2.99ex]{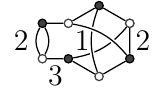}}}

\newcommand{\Aone}{\raisebox{-.30\height}{\hspace{-2pt}\includegraphics[height=3.05ex]{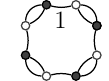}}}

\newcommand{\Aneutr}{\raisebox{-.30\height}{\hspace{-2pt}\includegraphics[height=3.05ex]{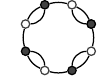}}}
\newcommand{\Sneutr}{\raisebox{-.30\height}{\hspace{0pt}\includegraphics[height=3.5ex]{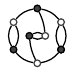}}}

\newcommand{\Rneutr}{\!\!\raisebox{-.172\height}{ \includegraphics[height=2.3ex]{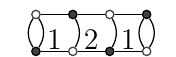}}}
\newcommand{\Triple}{\!\!\raisebox{-.172\height}{ \includegraphics[height=2.3ex]{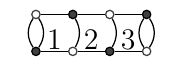}}}
\newcommand{\Mokka}{\raisebox{-.38\height}{ \includegraphics[height=3.5ex]{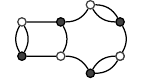}}}

\newcommand{\Mokkadt}{\raisebox{-.38\height}{ \includegraphics[height=3.5ex]{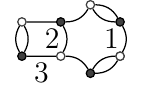}}}

\newcommand{\Mokkatd}{\raisebox{-.38\height}{ \includegraphics[height=3.5ex]{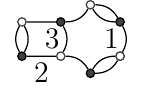}}}

\newcommand{\Eone}{\raisebox{-.48\height}{\hspace{.3pt}\includegraphics[height=3.5ex]{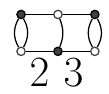}}\hspace{.5pt}}
\newcommand{\Etwo}{\raisebox{-.48\height}{\hspace{.3pt}\includegraphics[height=3.5ex]{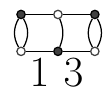}}\hspace{.5pt}}
\newcommand{\Ethree}{\raisebox{-.48\height}{\hspace{.3pt}\includegraphics[height=3.5ex]{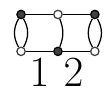}}\hspace{.5pt}}

\newcommand{\conn}{^{\text{\tiny conn.}}}

\newcommand{\logo}[4]{\raisebox{-.#4\height}{\includegraphics[height=#3 ex]{Logo#1_#2.pdf}}}

\makeatletter
\newcommand{\vast}{\bBigg@{4.030}}
\newcommand{\Vast}{\bBigg@{12.30}}
\makeatother
\newenvironment{Figure}
  {\par\medskip\noindent\minipage{\linewidth}}
  {\endminipage\par\medskip}
\usepackage{tabularx}

\def\[#1\]{%
  \begin{align}#1%
  \end{align}%
}

\definecolor{azulESI}{HTML}{1266AE}

\definecolor{AZULESI}{HTML}{1266AE}
\usepackage{enumerate}

\usepackage{nicematrix}
\usepackage{soul}
\usepackage[utf8]{inputenc}

\makeatletter
\ifcase \@ptsize \relax
\newcommand{\nano}{\@setfontsize\miniscule{3.5}{4.5}}
\or
\newcommand{\nano}{\@setfontsize\miniscule{4.5}{5.5}}%
\or
\newcommand{\nano}{\@setfontsize\miniscule{4.5}{5.5}}%
\fi
\makeatother

\usepackage{kbordermatrix}
\usepackage{blkarray}

\newcommand{\balita}{\raisebox{-0pt}{\text{\nano$\bullet$\hspace{.75pt}}}}

\usepackage{tikz}
\usetikzlibrary{automata,arrows,topaths}
\usepackage{caption}

\usepackage[bookmarks=false]{hyperref}
\hypersetup{
         colorlinks   = true,
         citecolor   =azulESI,
        urlcolor = azulESI
}
\hypersetup{linkcolor=azulESI}

\newcommand{\kthree}{\raisebox{-.2\height}{\hspace{1pt}\includegraphics[height=2.4ex]{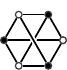}}\hspace{1pt}}
\newcommand{\kthreesumkthree}{\raisebox{-.327\height}{\hspace{1pt}\includegraphics[height=2.4ex]{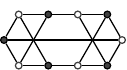}}\hspace{1pt}}
\newcommand{\Qone}{\raisebox{-.3\height}{\hspace{1pt}\includegraphics[height=2.4ex]{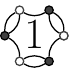}}}
\newcommand{\Qtwo}{\raisebox{-.3\height}{\hspace{1pt}\includegraphics[height=2.4ex]{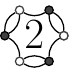}}}
\newcommand{\Qthree}{\raisebox{-.3\height}{\hspace{1pt}\includegraphics[height=2.4ex]{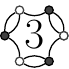}}}

\newcommand{\littlemelon}{\,\logo{2}{Melon}{2}{20}\,}
\newcommand{\littlemelonR}{\,\logo{2}{MelonR}{2}{20}\,}

\newcommand{\Cubon}{
    \raisebox{-.38\height}{\includegraphics[height=4ex]{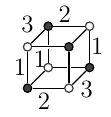}}
    }

\newcommand{\vone}{\raisebox{-.3\height}{\includegraphics[height=2.3ex]{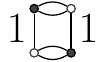}}}
\newcommand{\vtwo}{\raisebox{-.3\height}{\includegraphics[height=2.3ex]{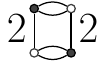}}}
\newcommand{\vthree}{\raisebox{-.3\height}{\includegraphics[height=2.3ex]{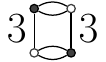}}}

\newcommand{\inv}{^{-1}}

\newcommand{\Vw}{V_{\circ}}
\newcommand{\Vb}{V_{\balita}}

\newcommand{\meas}{\dif T \wedge \dif \bar T }

\usepackage{float}
\usepackage{subcaption}
\usepackage{amsthm}
\numberwithin{equation}{section}
\newtheoremstyle{mytheoremstyle} 
    {10pt}                    
    {8pt}                    
    {\itshape}                   
    {}                           
    {\scshape}                   
    {.}                          
    {.5em}                       
    {}  

\makeatletter
\newcommand{\leqnomode}{\tagsleft@true}
\newcommand{\reqnomode}{\tagsleft@false}
\makeatother
\theoremstyle{mytheoremstyle}

\newtheorem{theorem}{Theorem}[section]

\newtheoremstyle{definition} 
    {8pt}                    
    {5pt}                    
    {}                   
    {}                           
    {\scshape}                   
    {.}                          
    {.5em}                       
    {}  

 \theoremstyle{definition}

\newtheorem{remark}[theorem]{Remark}

\newcommand{\N}{\mathbb{Z}_{>0}}
\newcommand{\Z}{\mathbb{Z}}

\newcommand{\C}{\mathbb{C}}

\newcommand{\runter}[1] {\raisebox{-.45\height}{#1}}

\newcommand{\runterdreiviert}[1] {\raisebox{-.22\height}{#1}}

\newcommand{\dif}{{\mathrm{d}}}

\newcommand{\uni}{U}

\newcommand{\ee}{\mathrm{e}}

\DeclareMathOperator{\Sym}{\mathrm{Sym}}

\DeclareMathOperator{\Tr}{Tr}

\newcommand{\hp}[1]{^{(#1)}}

\usepackage[includeheadfoot,margin=2cm]{geometry}
\usepackage[font=small,labelfont=bf,tableposition=top]{caption}
\usepackage{multicol,caption}

\tikzcdset{arrow style=tikz, diagrams={>={Stealth[round,length=4pt,width=4.95pt,inset=2.75pt]}}}

 \title[Bootstrapping Tensor Integrals]{Bootstrapping tensor integrals}
 \author[N. Pagliaroli, C. I. Perez-Sanchez and B. Smith ]{Nathan Pagliaroli,  Carlos I. P\'erez S\'anchez,  Brayden Smith}

\address{University of Waterloo, Department of Combinatorics and Optimization,\newline \indent
 200 University Ave West, Waterloo, Ontario, N2L 3G1, Canada}
  \email{npagliar@uwaterloo.ca}

  \address{University of Heidelberg, Institute for Theoretical Physics,\newline \indent
  Philosophenweg 19, 69120 Heidelberg, Germany,  EU
 }
  \email{perez.sanchez@protonmail.ch}

\address{McGill University, Department of Mathematics and Statistics,\newline \indent
  805 Sherbrooke St W, Montreal, Quebec, H3A 2A7 Canada}
  \email{brayden.smith@mail.mcgill.ca}
\usepackage{stackrel}

\makeatletter

\newcommand*\notocchapter[1]{%
  \if@openright\cleardoublepage\else\clearpage\fi
  \thispagestyle{empty}\global\@topnum\z@
  \@afterindenttrue
  \let\@secnumber\@empty
  \@makeschapterhead{#1}\@afterheading
}

\makeatother

\newcommand{\ErW}[1]{\mathbb{E}_g \left[ #1\right]}
\begin{document}
\begin{abstract}
This work proposes a bootstrapping with positivity methodology to
study random $U(N)^{D}$ invariant tensors in the large $N$ limit. As
has been done for $U(N)$ invariant random matrices, we combine the
Dyson-Schwinger equations and positivity constraints of moments to approximate the moments of such tensor models. As
examples, we bootstrap the quartic and two hexic rank three
tensor models. All models studied converge quickly, and for those which have known analytic
formulae, they converge to such solutions.  We conjecture new explicit formulae
for all moments of the rank three quartic model and support this
conjecture using bootstrapped results and explicit double-series computations with
\texttt{feyntensor}.
\end{abstract}
 \maketitle%

\section{Introduction}

Random tensors are a natural generalization of random matrices. As
random matrices are studied for their applications to two dimensional
Euclidean quantum gravity \cite{di19952d,stanford2020jt}, tensor
models have been proposed as models for higher dimensional theories
\cite{rivasseau2016random,Gurau:2009tw}. In particular, $U(N)^{D}$
invariant tensor integrals are the weighted formal generating
functions of $(D+1)$-coloured graphs, which can be considered discrete
Euclidean $D$ dimensional space-times (orientable piecewise linear
pseudomanifolds).  Many tensor integrals can also be related to matrix
integrals through intermediate field theory
\cite{lionni2019intermediate, Reiko_Intermediatefield}. Recently, just
as with random matrices, random tensors have found applications in
generalizing free probability theory \cite{collins2024free}.  Deeply
understanding the spectral properties for tensors
\cite{GurauWigner}, as well as the computation of eigenvalues and
their eigenvectors, are challenging problems. This is in particular
due to the plethora of definitions for tensor eigenvalues and their
computability being NP-hard \cite{ReikoNicolasSasakura_Eigenpairs}.

In general,  analytical formulae are scarce  for tensor
integrals of any rank, with  some notable exceptions such as in
\cite{gurau2014universality,nguyen2015analysis,lionni2019multi,bonzom2017counting}. However,
much is known about the critical behaviour and exponents of tensor
models. If the number of associated graphs grows sufficiently quickly, formal
matrix and tensor models are well-defined generating functions with
critical points. The moments and free energy of such models are often
algebraic, with a critical exponent called the string susceptibility
$\gamma$. When $\gamma = \frac{1}{2}$, the model's graphs are usually
in bijection with some sort of plane tree for which the continuum
limit is the Continuum Random Tree
\cite{gurau2014melons,aldous1991continuum}. If $\gamma =
-\frac{1}{2}$, the graphs are usually in bijection with decorated
plane trees and the associated continuum limit is usually the Brownian
map \cite{le2013uniqueness,miermont2013brownian}. In multi-tracial matrix models and
certain tensor models, other critical exponents are possible
\cite{lionni2019multi,korchemsky1992matrix}. When one has multiple
critical points, many more exponents are possible in multi-critical
regimes \cite{kazakov1989appearance}. For an overview of the critical
exponents of tensor models see Chapter Four of
\cite{lionni2018colored}. The critical points of tensor models can
also be studied via Renormalization Group techniques
\cite{castro2026towards}. In the recent work of two of the authors,
the critical exponents of multi-matrix models were estimated using
bootstrapping with positivity \cite{khalkhali2025multi}. We will see
in this paper that the critical points and exponents of various tensor
models appear in a similar manner.

A plethora of methods exist for studying single matrix models
\cite{deift2000orthogonal,eynard2016counting}. The lack of analytic
methods for multi-matrix models first motivated  Lin \cite{lin2020bootstraps}
to apply the method of bootstrapping with positivity to these matrix
integrals. To the authors' knowledge, the method of bootstrapping with positivity appeared first in lattice gauge theory
\cite{Kruczenski}, where the Dyson-Schwinger equations were combined
with positivity constraints on matrix moments. For matrix ensembles,
this combination leads to bounds and approximations of
their tracial moments
\cite{kazakov2022analytic,hessam2022bootstrapping,li2025bootstrapping}. This
method has also seen a flurry of use not just for random matrices, but also
in many other areas of theoretical physics
\cite{guo2025bootstrapping,cho2025bootstrapping,zeng2023feynman,khalkhali2025bootstrapping,perez2025loop},
particularly in matrix quantum mechanics
\cite{berenstein2023semidefinite,berenstein2024one,huang2025bootstrapping,li2024bootstrapping,sword2024quantum}.  Recent work has  seen success bootstrapping matrix models with different constraints than positivity \cite{maeta2026matrix,li2025analytic}, as well as the reconstruction of the eigenvalue distribution of random matrices from bootstrapped bounds \cite{kovavcik2025eigenvalue}.  Just
as with matrix integrals, the situation is ripe for studying tensor
integrals via bootstrapping with positivity. 
(In fact, while we were concluding this work,  bootstrapping with positivity was independently and promisingly applied for bootstrapping matrix and tensor models at finite size \cite{toriumi2026}.)
Tensor integrals have
both a set of infinite recursive equations for moments (Dyson-Schwinger equations \cite{GurauVirasoro,Krajewski:2012dq,fullward}) as well as positivity constraints on
moments that can be used for such a scheme. 

\begin{figure}
\huge \raisebox{-.4\height}{\includegraphics[width=.15\textwidth]{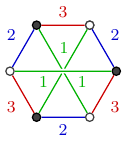}} $  \hookrightarrow $
\raisebox{-.4\height}{\includegraphics[width=.34\textwidth]{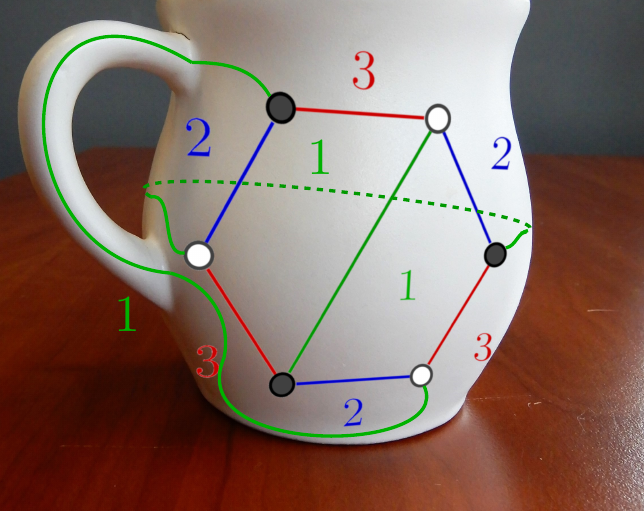}} \normalsize
\caption{The graph corresponding to the hexic tensor $U(N)^3$-invariant on the left is embedded in
a genus-$1$ surface (minimal genus guaranteeing a planar embedding).\label{fig:mug} }
\end{figure}

In this paper, all tensor integrals will involve unitary invariants
motivated by Gur\u au's initial tensor models.  Here, for tensors with
$D$-indices, such unitary invariants are in bijection with bipartite
regularly $D$-edge-coloured graphs (or $D$-coloured graphs for
conciseness), which have been enumerated in
\cite{counting_invariants,OEIS}.
The dual graphs are interpreted as
gluings of $(D-1)$-simplices (associated to the vertices), where the
colouration of the edges indicates the manner in which one glues the
facets. Thereby, the bipartiteness yields a consistent way to orient
the resulting space. Thus, a $3$-coloured graph $G$  triangulates a compact, boundaryless oriented surface which is determined
by its genus $g$ (Classification Theorem). This is given by
\[ 2-2g = \sum_{i=0,1,2} (-1)^i \#\{i\text{-coloured connected subgraphs of } G\},\] and
$g$ coincides with the lowest-genus surface that
allows a planar embedding of that graph (see Fig. \ref{fig:mug}). Deeper
enumerative aspects of such graphs and their gluings are for instance
addressed in \cite{bonzom2017coloured,bonzom2017counting}.

In view of novel results regarding the
non-factorization \cite{gurau2025nonfactor} of arbitrary multi-trace invariants (corresponding to
disconnected coloured  graphs), we present a
conjecture that could be of interest. We use \texttt{feyntensor} \cite{feyntensor} to  obtain
expressions for the double-series (in the coupling constant and in $1/N$) which
suggest that, for integrals over tensors with three indices, the genus (and not whether it is melonic) plays the decisive role in a non-factorization of tensor invariants (henceforth `invariants') in the following sense:
It had once already been conjectured that
the leading order of any (single-trace properly scaled, see e.g. \cite[Eq. 3.19]{uncolouring}) observable $B$ factors in the large-$N$ limit as a constant $C_B$
times a power of the two-point function $m_2$,
thus as  $C_B \cdot (m_2)^{\deg B/2}$, which in turn is solved by the Fuss-Catalan numbers
that appear also in vector models \cite[App. IV]{rivasseau2017loopvertexp}.
According to another conjecture we present below, a necessary condition for
the expectation of a trace not to factor as above
as a power of the two-point function is the presence
of non-planar---which is not the same as non-melonic---graphs in the integrand (but
this condition is not sufficient). Furthermore, as a corollary of our conjecture we obtain explicit formulae for the expected value of all rank three tracial invariants for the quartic tensor model of rank three in the large $N$ limit. This conjecture would generalize the known formulae for the second and fourth moment of the quartic model \cite{nguyen2015analysis}.

In Section \ref{sec:TM}, we provide the necessary background on
tensor models and their coloured graphs, with a focus on rank
three. In Section \ref{sec:Bootstr}, we discuss how to derive the
Dyson-Schwinger equations and the positivity constraints on
moments. We then discuss the general algorithm for bootstrapping. In
Section \ref{sec:results} we bootstrap several rank 3 tensor models
and compare our results to known solutions. In Section
\ref{sec:conjecture} we present evidence from series expansions and
bootstraps for a conjecture on the moments of the quartic model.  In
Section \ref{sec:summary}, we summarize our results and discuss future
work. In Appendix \ref{app:scaling} we outline the
implementation of graph operations. Then in Appendix \ref{app:implementation}, we discuss our algorithm for computing the
scaling factor of invariants. In Appendix \ref{app:expansions}, some series expansions of cumulants are given. Lastly, in Appendix \ref{app:SDElist}, we give more examples of
the Dyson-Schwinger equations of the quartic model studied.

\section{Tensor models}\label{sec:TM}
A way of constructing actions in physics is to specify invariance
under a Lie group action. For instance, if the unitary group acts on
the fundamental representation, an action is spanned by
$\uni(N)$-invariants, namely powers of $\sum_i \bar w_i w_i$ 
for a vector $w=\sum_i w_i e_i \in \C^N$, where $\{e_1,\ldots, e_n\}$ is the
standard orthonormal basis of $\C^N$.  The resulting theories are known as
vector models, and one can similarly define matrix models in terms of
such invariance, leading to traces of the object in question.  Even
more generally let $T$ be a complex covariant tensor of rank $D$, i.e. a
multi-linear map on the tensor product $\C^N\otimes\cdots
\otimes \C^N$. For a fixed basis, denote its entries
$T_{a_{1},a_{2},...,a_{D}}$ and $\overline{T}_{a_{1},a_{2},...,a_{D}}$
for the entries of the complex conjugate, where indices range from one to $N$. We
say that a polynomial in the entries of $T$ and $\overline{T}$ is an \textit{invariant} if it is invariant under the action of
$U(N)^{\otimes D}$:
	\begin{align*}
			T_{a_{1},a_{2},...,a_{D}} &\rightarrow\sum_{b_1, \cdots, b_D} U_{a_1, b_1}^{(1)} \cdots U_{a_D, b_D}^{(D)} T_{b_1, \cdots, b_D}, \\
			\overline{T}_{a_{1},a_{2},...,a_{D}} & \rightarrow \sum_{c_1 ,..., c_D} \bar{U}_{a_1 c_1}^{(1)} \cdots \bar{U}_{a_D c_D}^{(D)} \overline{T}_{c_1 ,\ldots,c_D}.
		\end{align*}

The polynomials that are invariant under this action are such that the $i$th index of $T$ is summed with the $i$th index of $\overline{T}$. An invariant, denoted by $\Tr_{G} (T,\overline{T})$, has a unique graph representation $G$ where:
		\begin{itemize}
			\item $T$ indices are represented as white vertices.
			\item $\overline{T}$ indices are represented as black vertices.
			\item The sum over the $i$th index shared by $T$ and $\overline{T}$ is represented by a line of colour $i$.
	   \end{itemize}
For example, constructing $\uni (N)^D$-invariants
out of \[T = \sum_{a_1,a_2,\ldots, a_D=1}^N
T_{a_1,a_2,\ldots, a_D} e_{a_1} \otimes e_{a_2} \otimes  \cdots \otimes e_{a_D}
\in  (\C^N )^{\otimes D} \] requires a choice of the many\footnote{Cf. \cite{counting_invariants} for the generating function of these numbers.}
index contractions at a fixed degree in $T$ (which by invariance, has to coincide with that of $\overline{T}$).
For instance, if $D=3$, in a quartic invariant
there are four ways of contracting (from this point on, we use Einstein's summation for repeated indices) the twelve indices:%
\begin{subequations}%
 \begin{align}%
  T_{a_1,a_2,a_3}   \bar  T_{b_1,a_2,a_3} T_{b_1,b_2,b_3}    \bar  T_{a_1,b_2,b_3} &\leftrightarrow
 \vone \\
 T_{a_1,a_2,a_3}   \bar T_{a_1,b_2,a_3} T_{b_1,b_2,b_3}    \bar  T_{b_1,a_2,b_3} &\leftrightarrow
 \vtwo \\
T_{a_1,a_2,a_3}   \bar  T_{a_1,a_2,b_3} T_{b_1,b_2,b_3}    \bar  T_{b_1,b_2,a_3} & \leftrightarrow
 \vthree \\
  T_{a_1,a_2,a_3}   \bar  T_{a_1,a_2,a_3} T_{b_1,b_2,b_3}    \bar  T_{b_1,b_2,b_3} & \leftrightarrow\,
\underset{\littlemelon}{ \littlemelonR} \vspace{-1ex}
\end{align}%
\end{subequations}
We shall write $\Tr_B(T,\bar T)$ for the invariants resulting by this correspondence. 
Connectedness of invariants will be interpreted
in terms of their respective graphs. Thus, only the last invariant in the list
above is disconnected.

In general, denote the rank $D$ melon invariant as
$$T\cdot \overline{T} = 
T_{a_{1},...,a_{D}} \overline{T}_{a_{1},...,a_{D}}.$$
Then we can define the Gau\ss ian measure on the complex vector space of tensors
		\begin{equation}\label{measureT}
			\dif \mu_{N}(T, \bar{T})=\exp \left(-N^{D-1} T \cdot \bar{T}\right) \dif \bar T  \dif T,
		\end{equation}
        where
        $$ \dif \bar T \dif T=\prod_{a_1, \ldots, a_D} \bigg(\frac{N^{D-1}}{2  \pi i } \bigg)\dif \bar T_{a_1, \ldots,  a_D}\wedge \dif {T}_{a_1, \ldots,  a_D} .$$
That the quantity in brackets is the correct normalization constant can be verified by integrating
        over the $2N^D$ real independent degrees of freedom. For a fixed tensor component $T_{a_1\ldots a_D}$ ($a_i \in \{1,\ldots,N\}$),
        $\dif \bar T_{a_1,\ldots, a_D} \wedge \dif T_{a_1,\ldots, a_D} = 2i \dif x_{a_1\ldots a_D} \dif y_{a_1\ldots a_D}$ in terms of the real ($x_{a_1\ldots a_D}$) and imaginary ($y_{a_1\ldots a_D}$) components of $T_{a_1, \ldots,  a_D} $. One can compare the previous measure with the
        known normalization of Hermitian matrix integrals;
        if one requires that $ \int  \exp(-N \Tr M^2)\dif M  =  1$
        then the Lebesgue measure $\dif M$ over the $N^2$
        real components of the Hermitian matrix should be
        divided by ${2}^{N/2} {(\pi/N)}^{N^2/2}.$ (The formula does not reduce to the matrix normalization constant when $D=2$, but this should not alarm us, since our tensor is complex without symmetries.)

A tensor model is a formal integral of the form\footnote{
Later we will generalize allowing a $N^{s_{j}}$ scaling factor for the interaction $C_j$ and solve
 for these.}
		\begin{equation*}
			Z(N,g)=\int \exp \left(-N^{D-1} \sum_{j}   g_{j} \operatorname{Tr}_{C_{j}}(T, \bar{T})\right) \dif \mu_{N}(T, \bar{T}),
		\end{equation*} and
$g=\{g_j\}$ is a list of real couplings corresponding to a finite set $ \{C_j\}_j$
of $D$-coloured graphs, which in their role of interactions are traditionally called \textit{bubbles}. By formal integral, we mean a formal summation formed by
series expanding the exponential and swapping the order of integration
and summation. The result is a weighted sum of Gau\ss ian moments. By
Wick contracting edges of invariants with the Gau\ss ian measure one can
obtain an expansion in $1/N$
	\begin{equation*}
	\log Z =\sum_{G}N^{D- \frac{2}{(D-1)!} \omega(G)}\frac{\prod_{j}g_{j}^{ n_{j}(G)}}{\text{Sym}(G)},
	\end{equation*}
	where
    \begin{itemize}
        \item the sum is over the set of connected coloured graphs $G$ in $D+1$ colours,
        that can be glued from bubbles $ \{C_j\}_j$ (hence the additional colour accounting for the propagator)
        \item $\text{Sym}(G)$ is a symmetry factor \cite{bonzom2016large},
         \item $n_{i}(G)$ denotes the number of bubbles $C_i$ used in gluing $G$,
        \item and $\omega(G)$ is Gur\u au's degree \cite{gurau2016regular} characterized by
        \begin{equation*}
           D-\frac{2}{(D-1)!} \omega(G) := \# \text{faces of $G$}  - \frac 12 \binom{D}{2} \# \text{vertices of $G$} \leq  D
            \end{equation*}
        where a face of $G$ is a $2$-coloured connected subgraph of $G$. For the $1/N$-expansion,  it is important that
        $ \omega (G) \geq 0$ holds.
        \end{itemize}

When $\omega(G)=0$, then $G$ is glued from \textit{melonic bubbles}.
Such an expansion also exists for expectation value of
invariants, in which melonic bubbles also dominate in the large $N$
limit. A $D$ dipole graph of colour $c$ is the graph with two vertices
connected by the $D$ edges of all colours except $c$ and half-edges of
colour $c$ incident to both vertices. A \textit{melonic bubble} is a
graph that can be obtained by iteratively inserting dipoles onto edges
of colour $c$ of the fundamental melon.

In this paper, we will outline a general process for bootstrapping
tensor models of any rank, but we restrict computations to rank
three. A rank $D$ tensor model of the form
\begin{align}\label{model_unscaled}
S_{g}(T,\bar T)= T\cdot \bar T +\sum_{j} N^{s_j}g_{j}
\operatorname{Tr}_{C_{j}}(T, \bar{T})\end{align} can be then specified by a
set $\{C_j\}_{j=1}^n$ of $D$-coloured graphs.  This
abbreviates that graphs are vertex-bipartite (vertices are either
black or white), that at each vertex precisely $D$ edges, each having a
different colour out of $\{1,2,\dots ,D\}$, and each edge connects only
vertices of different vertex-type, i.e. a white vertex with a black
one. Given the action $S_{g}$ as in
\eqref{model_unscaled}, the expectation values for each unitary
invariant and corresponding $D$-colour graph $G$ are
\[ \ErW{ \operatorname{Tr}_G(T, \bar{T})} & =\frac1{Z(N,g)} \int
\operatorname{Tr}_G(T, \bar{T}) \ee^{-S_g( T,\bar T )} \,\dif T \dif \bar T
\\ &= \frac 1 {Z(N,g)}\int \operatorname{Tr} _G(T,\bar T) \ee^{-\sum_{j} N^{s_j}g_{j}
\operatorname{Tr}_{C_{j}}(T, \bar{T}) }   {\dif \mu_N(T,\bar T)} \label{expectation_vals}. \]

With a focus on rank three tensor models, it is the aim of this article to determine for which choices of the
action $S$ and couplings the model does or does not exist as a
solution to the Dyson-Schwinger equations that satisfy essential
positivity constraints. This combination produces bounds with remarkable computational efficiency,
determining the approximate solution. By `solving' we mean to determine
the expectation values as a function of the couplings, at least in the
large-$N$ limit.

\section{Bootstrapping with positivity} \label{sec:Bootstr}

\subsection{Tensor positivity constraints}\label{sec:2of3}
Given the straightforward relation $  P \cdot  \bar P  \geq 0$,
 which holds for any complex valued tensor $P_{x_1,x_2,x_3}$,
 we insert the choice $P_{x_1,x_2,x_3}= \partial_{x_1,x_2,x_3} \sum_{j=1,\ldots, k} z_j B_j$
 where $B_j \in \C [T,\bar T]$ is a unitary invariant
 and  $\partial_{x_1,x_2,x_3} = \partial / \partial T_{x_1,x_2,x_3}$.
 Here, $P=\sum_{j=1,\ldots, k} z_j\partial \maketrace{B_j}$ is used for the sake of notation.
 Similarly, if $\bar \partial_{x_1,x_2,x_3} = \partial / \partial \bar T_{x_1,x_2,x_3}$,
 then $\bar P = \sum_{j=1,\ldots, k} \bar  z_j   \bar \partial  \maketrace{ B_j }$, so
 \[ 0 \leq \sum_{i,j=1}^k z_i  (\partial  \maketrace{B_i }\cdot  \bar \partial  \maketrace{B_j}  )\bar z_j\,,
  \]
 for any $z=(z_1,\ldots,z_k)\in \C^k$. The independence of this inequality on the vector $z \in \C^k$
 implies the positivity
 of the matrix $\mathcal W=(\mathcal W_{i,j})_{i,j=1,\ldots,k} $ whose entries are given by the expectation values
 \[
 \mathcal W_{i,j}(g)= \ErW {   \partial \maketrace{B_i} \cdot \bar \partial  \maketrace{ B_j}    } \in \C [\![g]\!]\, ,
 \]
 which is obviously self-adjoint for real $g$.

 We now describe how to graphically obtain the argument of the matrix elements. We restrict this discussion to 3-regular graphs for its relevance to our investigations.
 Given a   $3$-coloured graph $B$ let $\Vw( B) $ denote it's set of white vertices
   and  $\Vb( B) $ that of its black vertices. Then $\partial \maketrace B$  (resp. $\bar\partial \maketrace B$)
   is the sum of excisions of all white-vertex (resp. all black-vertex) of $B$.
   When we sum over indices to form $\partial \maketrace {B_i} \cdot \bar \partial \maketrace {B_j}$,
   we connect the broken half-edges respecting the colouration. One defines
  \[ A   \tensor[_u]{\star}{_{v}}     B :=
   \big [ (  A   \dot\cup   B) \setminus \{u,v\}\big]_{\text{welded}}\,,
  \]
 where $  A\dot\cup   B$ denotes the disjoint union of the coloured graphs $A$ and $B$ (that is,
 $  A\dot\cup A$ is two copies of $  A$). The extraction of the vertices $u$ and $v$ from the
 graph  $  A\dot\cup   B$ leaves three coloured edges in $A$ and three in $B$
 that are matched respecting their colouring (this is what `welded' means).
 For instance $\vone   \tensor[_u]{\star}{_{v}}    \vtwo = \Ethree$, where
 $u$ is any black vertex of $\vone$ and $v$ any white vertex of $\vtwo$.

 We conclude that
 \[
 \mathcal W_{i,j}(g) = \sum_{\substack{ u\in \Vw( B_i) \\ v\in\Vb( B_j) }} \ErW{  \maketrace {
 (B_i)   \tensor[_u]{\star}{_{v}}     (B_j) } \label{defW}
 }\geq 0.
 \]
By positivity, taking principal minors of $\mathcal{W}$ of our matrix, denoted $\mathcal{W}_{I}$, where $I \subset \mathbb{N}_{+}$, must also be positive semi-definite, i.e. all eigenvalues are greater than or equal to zero. We denote the set $\{1,2,3,...,n\}$ by $[n]$. By considering the principal minors eigenvalues, we have positivity constraints on our moments. 

%
%
%
%
Noticing that a permutation matrix is unitary, which will not affect the positivity
condition, we chose the following order of the list of
observables\footnote{An abbreviated notation is used, where
only a distinguished edge is coloured. E.g. in $\Eone$ uniform edge
colouration implies that the two vertical edges from left to right have
colours $\{1,3\}$, the middle vertical one $\{1\}$ and the rightmost
vertical edges $\{1,2\}$. Also in $\littlemelon$ and $ \kthree$ no
numbers are displayed, since any (legal) colouration will lead to the
same graph.}
\begin{align}
  \{ \littlemelon,\, \tfrac 12 \vone\,, &\tfrac 12 \vtwo\,, \tfrac 12 \vthree\,, \tfrac 13  \kthree\,,  \notag \\
 &\tfrac 13 \Qone\,, \tfrac 13 \Qtwo\,,\tfrac 13 \Qthree\,, \Eone, \Etwo,\Ethree, \ldots \} \label{list}.
\end{align}
Using this list, we compute the matrix elements $(\partial B_i \cdot
\bar \partial B_j )$ first. In order to obtain their monic monomials,
we divide by the symmetry factor of the graph in \eqref{list} (for
graphs with more than two vertices and non-trivial automorphism this
is not always possible).  This leads to the following matrix of invariants
\[
 \kbordermatrix{
 \, \downarrow \,\, \cdot \,\rightarrow\,  &  \vphantom{\Big\{}\bar\partial \littlemelon&\frac 12 \bar\partial \vone &\frac 12 \bar\partial  \vtwo & \bar \partial  \vthree  & \frac13\bar\partial \kthree & \cdots \cr
                \partial \littlemelon&  \littlemelon& \vone  &  \vtwo  &  \vthree & \kthree  & \cdots \cr
               \frac12 \partial \vone & \vphantom{\Big\{}  \vone   &  \Qone &  \Ethree & \Etwo & \Xu & \cdots \cr
               \frac12 \partial \vtwo & \vphantom{\Big\{} \vtwo  & \Ethree & \Qtwo  & \Eone & \Xd & \cdots \cr
                \frac12\partial \vthree &\vphantom{\Big\{}  \vthree  & \Etwo       &  \Eone&  \Qthree & \Xt   & \cdots \cr
                  \frac13\partial \kthree & \vphantom{\Big\{} \kthree  &  \Xu       &  \Xd &  \Xt & \kthreesumkthree  & \cdots \cr
                \vdots & \vdots  & \vdots & \vdots  & \vdots  &  \vdots & \ddots
                }.
\]
The arrows in the upper corner point towards the arguments that the
product $\cdot$ takes, and $\mathcal W$'s entries are the expectation values of the previous matrix' entries.

\subsection{Coloured matrix positivity constraints}
For the hexic model studied below, we only use  the constraints from the the positive semi-definiteness of 
$\mathcal W$, given by in Eq. \eqref{defW}. However, for the quartic model, these constraints only provided a lower bound on $m_{2}$, all be it very precise. We therefore require additional bounds, and apply those obtained in \cite{toriumi2026additional}, which provide us with a flush upper bound for $m_{2}$ of the quartic model.

For a 3-coloured graph $B$ with an edge $e$, the \textit{colour matrix graph}  $B\setminus e$ is the graph obtained from $B$ after replacing $e$ by two half-edges. More specifically, the expression corresponding
to $B \setminus e$ in terms of $T$ and $\bar T$  is just the invariant $B$
with the missing Kronecker delta (that corresponded to $e$) and omitting the respective sum over those two indices.
For instance,
picking any of the colour-$1$ edges from each graph of the list
$  \{ \littlemelon,\, \vone\,,  \vtwo\,, \vthree\,, \ldots \} $ 
one obtains the basis 
\[  \notag
\{&
T_{i_1,a_2,a_3}
\bar T_{j_1,a_2,a_3},
T_{i_1,a_2,a_3}
\bar T_{b_1,a_2,a_3}
T_{b_1,b_2,b_3}
\bar T_{j_1,b_2,b_3},
\\
&T_{i_1,a_2,a_3}
\bar T_{j_1,a_2,b_3}
T_{b_1,b_2,b_3}
\bar T_{j_1,b_2,a_3},
T_{i_1,a_2,a_3}
\bar T_{j_1,b_2,a_3}
T_{b_1,b_2,b_3}
\bar T_{j_1,a_2,b_3},
\ldots \notag
\}
\]
Notice that the  indices $i_1,j_1$ are free and the rest summed over; thus the elements of this list are the entries of a matrix, which being
all at the tensor's  first  entry explains  the name  \textit{coloured matrices}. These matrix elements lead to a different set of positivity constraints,
which one can derive in the same manner as in the previous section, but with the Hilbert-Schmidt inner product of matrices instead of the product on $(\C^N)^{\otimes 3}$. In fact, this allows to add  $\delta_{i_1,j_1}$ to the basis
above and obtain the following positive semi-definite matrix of moments
\[\mathcal M= \mathbb E_g \label{AddConstrMatrix}
\begin{bmatrix}
 N & \littlemelon & \vone & \vtwo  & \vthree   & \cdots \\[.85ex]
\littlemelon & \vone & \Qone & \Ethree & \Etwo  & \cdots  \\[.85ex]
 \vone &  \Qone & \Aone    & \Mokkatd & \Mokkadt  & \cdots  \\[.85ex]
\vtwo  &  \Ethree & \Mokkatd & \runterdreiviert{\includegraphics[height=2.4ex]{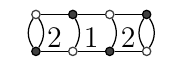}} & \runterdreiviert{\includegraphics[height=2.4ex]{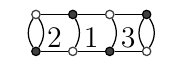}} & \cdots
\\[.85ex]
\vthree & \Etwo &  \Mokkadt  & \runterdreiviert{\includegraphics[height=2.4ex]{Item8_W3.pdf}} & \runterdreiviert{\includegraphics[height=2.4ex]{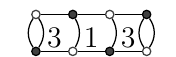}} & \cdots \\
                \vdots & \vdots  & \vdots & \vdots  & \vdots  & \ddots
 \end{bmatrix}
\]
where the expectation value is taken entry-wise.
This  matrix $\mathcal M$ has obvious generalizations for
other bases, and although it depends on the basis choice, we opt for
a simple notation here.
Note that the matrices $\mathcal{W}$ and $\mathcal{M}$ are both analogous to the Hankel matrix in the Hamburger moment problem. However, in our case, observe that this is not a matrix of the Hankel form:
in particular, the anti-diagonal is not constant.

\begin{figure}
\includegraphics[width=.6\textwidth]{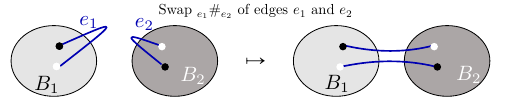}
\caption{Swap at two edges $e_1$ and $e_2$ of the same colour (here in blue),
belonging to two graphs $B_1$ and $B_2$, respectively. All non-depicted edges are not modified.\label{fig:swap}}
\end{figure}

We can re-express the coloured
matrix $\mathcal M$ of \cite{toriumi2026additional} very succinctly
in terms of the swap $ B_1 \#  B_2$ of graphs \cite{fullward}.
For graphs $B_1$ and $B_2$ with the same number of colours, this is
defined in Figure \ref{fig:swap}. For a given basis $\{B_a\}_{a=1}^n$ (for some $n$)
of observables and (say) 1-coloured edges $e_a$ belonging to $B_a$ for each $a$,
\[
\mathcal M_{a,b} := \mathbb E_g [ (B_a)\,{} _{e_a}\#_{e_b} \,(B_b) ]  \qquad a,b=1,\ldots,n . \label{connsum}
\]
and $\mathcal M_{a,0}:=  \mathbb E_g [ B_a ] =: \mathcal M_{0,a}$ for any $a=1,\ldots,n$, and $\mathcal M_{0,0}:=N$.
Observe how Eq. \eqref{AddConstrMatrix} is obtained from the basis
$  \{ \littlemelon,\, \vone\,,  \vtwo\,, \vthree\,, \ldots \}$ (with selected $1$-coloured edges)
using coloured graphs
by Prescription \eqref{connsum}, without reference to matrices.

As a caveat, the generalization of $\mathcal M$ for bases that include non-melonic 
observables is slightly subtle in view of the needed scalings. Since non-melonic observables are beyond our scope, we refer to \cite{RedundantMoments}.

\subsection{Dyson-Schwinger equations}\label{sec:1of3}
Fixing $N \in \N$, we first consider a generic model
\begin{align}\label{model}
S_{N, g, s}(T,\bar T)=N^{s_0} T\cdot \bar T + \sum\limits_{j=1}^n
N^{s_j} g_j
{{C}_j}\end{align}
where the scaling factors $ s=(s_0,s_1,\ldots, s_n)\in \mathbb R^{n+1}$ have to be determined.
Let us start with the identity
\[
\int  \partial \cdot  \big  [ \ee^{-S_{N, g, s}(T,\bar T) } \bar \partial \maketrace{B}   \big]\,  \meas  = 0,\label{eq:stoke's}
\]
where $B $ is a connected graph. Furthermore, the dot notation
has been extended to differential operators. For example when $D=3$: $\partial \cdot   \bar \partial P =  \sum_{x_1,x_2,x_3=1}^N \partial_{x_1,x_2,x_3} \bar \partial_{x_1,x_2,x_3} P(T,\bar T)$
for $P$ a function of $T$ and $\bar T$; a similar
sum abbreviates $\partial P\, \cdot \bar\partial Q =\partial_{x_1,x_2,x_3} P(T,\bar T)  \bar {\partial}_{x_1,x_2,x_3} P(T,\bar T)  $. The operator $\partial \,\cdot $ is the divergence,
a top-degree differential operator in $(\C^N)^{\otimes D}$. The identity above can be
re-expressed via integration by parts as
$\ErW{\partial \cdot \bar\partial \maketrace B }=\ErW{\partial \maketrace B\cdot \bar\partial S_{N, g, s}(T,\bar T) }$,
which, when expanded, reads
\begin{align}
 N^{s_0}&\sum_{v \in \Vw(B) }  \ErW{ \maketrace B } + \sum\limits_{j=1}^n
 g_{j}  N^{s_j}\sum\limits_{\substack{ w \in  \Vb(C_j)  \\ v \in  \Vw(B)  } }
 \ErW{  \maketrace  {B   \tensor[_v]{\star}{_{w}}   C_j} } \nonumber \\
& =    \sum\limits_{ \substack{ u \in  \Vb(B)  \\ v \in  \Vw(B)  }}
 N^{\# E  ( u, v) }
 \ErW{ \maketrace  { (B \setminus \{u,v\})\big|_{\text{\tiny welded} } }}\,, \label{DSE}
\end{align}
where for $u \in \Vw(B) $ a white vertex of $B$ and  $v \in \Vb(B)$ a black one. The
    graph $(B \setminus \{u,v\})\big|_{\text{\tiny welded}}$
  results from removing $v$ and  $u$ from $B$
 and subsequently gluing the `broken' half-edges that were either attached to $u$ or to $v$
but not to both (as $E(u,v)$,  the set of edges connecting the vertices $u$ and $v$,
accounts for those detached from both vertices),  respecting their colouring.
The correlators of the model \eqref{model}
thus obey these \textit{Dyson-Schwinger equations} (DSE).

These were first derived by Gur\u au in a context of a generalization of the Virasoro constraints
from matrix models to tensor models. They were
thus expressed as differential operators annihilating the partition
function; from them, a subset yields the Witt algebra
\cite{GurauVirasoro}. The DSE can be interpreted as in Figure
\ref{fig:DSE} in terms of the simplicial gluing $\Delta(B)$ that the
graph $B$ yields.

 \begin{Figure}\centering
 \includegraphics[width=.69\textwidth]{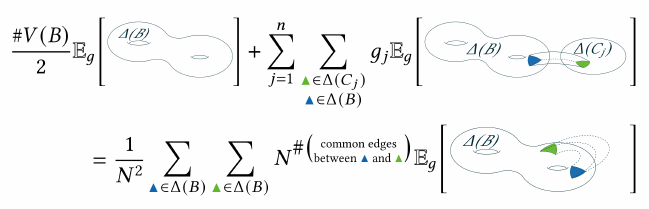}
 \captionof{figure}{The geometric interpretation of the DSE's (here with  $s=0$ on the L.H.S.)
 and $B$ connected\label{fig:DSE}. Here the blue triangles
 correspond to the black vertices and the green ones to the white vertices of the
 coloured graphs from which one obtains these triangulations. This figure is inspired by \cite[Eq. 5]{Krajewski:2012dq}.}
 \end{Figure}

For completeness, in general\footnote{The author thanks R\u azvan Gur\u au 
for comments that led to search for the present scaling 
of the moments (cumulants) and the vertices in the action.},  the moment of a graph $B$
where $\# \pi_0(B) $ denotes the connected components
remains finite and is of order $N^0$ if we set
\[
m_B:= \frac{1}{N^{\# \pi_0 (B)  -t(B) }  } \ErW{\maketrace B},
\]
where  $t(B)=0$ if $B$ is melonic,
and else $B$ is constructed as in Appendix \ref{app:scaling}, which nevertheless \textit{is valid only for melonic models} (i.e. all $C_j$'s are melonic, and else the algorithm to find $t$ is not known,
to the best of our knowledge). For instance, the graphs in \eqref{mkthreethree} and  \eqref{meightX} have a $t$-number equal to 1,
so they scale as $N^{1-1}=1$,
while the graph in \eqref{m22} with two melonic connected components, has a
scaling equal to $1/N^{2}$.

\begin{remark}
    Note that, whether one considers the perturbative definition of the partition function, i.e., as the weighted generating function of coloured graphs, or as an honest convergent integral, the identity \eqref{eq:stoke's} and the subsequent DSE for these models are the same. Additionally, in either case, the positivity constraints are also satisfied. Hence, any solution found via the DSE and these positivity constraints could belong to the formal integral or convergent integral interpretation of the model. In particular, if the solution is unique, which seems to be the case for all the models studied in this paper, then we can conclude that in the large $N$ limit both the formal and convergent models coincide. An analogous phenomenon is the case with Hermitian matrix models \cite{eynard2011formal}. There exist examples where a formal and corresponding convergent matrix models do  not have the same solution, and it would be interesting to find such a case for tensor models.
\end{remark}

\subsection{The general algorithm}\label{sec:3of3}

Given the DSE in the large $N$ limit, after appropriate scaling,
combined with the positivity constraints of the symmetric Hankel-like
matrix, we have a linear optimization problem with non-linear
inequality constraints. To make this problem computable, one can
truncate the choice of DSE as well as the basis of invariants for the
symmetric matrix. This is in contrast to the same method for matrix
models \cite{li2024bootstrapping} in which we have by default a
non-linear optimization problem. Although, it is worth noting that the
relaxation bootstrap method for matrix models is linear
\cite{kazakov2022analytic}. Additionally, note that
the DSE on their own are a very under-determined system of equations,
while the Dyson-Schwinger equations of many matrix models only
require a finite number of moments to determine all moments
recursively.

A \textit{Python} script generates all the invariants, their DSE, and
the positivity constraints from the symmetric matrix up to specified
cutoffs. For details on how the necessary graph operations are
implemented see Appendix \ref{app:implementation}. The \textit{Matlab}
non-linear optimization problem solver \textit{fminsdp} is then used to find possible solutions for a
given set of couplings. The function to be
minimized is set to $\sum_{i=1, \text{odd}} \cos(\theta+\pi/i) x_{i} +
\sum_{i=1, \text{even}}\sin(\theta+\pi/i) x_{i} $, where $(x_{i})$ is
the list of expectation values of invariants that appear in the DSE
and $\theta\in [-\pi,\pi]$. For a fixed value of the coupling
constants, the default interior-point algorithm is then used to find
points that satisfy the DSE and positivity constraints starting from
an initial guess for the expectation value of each invariant, for
various values of $\theta$. Repeating this process for various values
of the couplings allows one to generate plots of regions containing
the solution(s). As one might expect, increasing the number of DSE and
the size of the symmetric matrices both seem to decrease the area of
the solutions space. It is worth noting that this convergence seems
much slower than in matrix models studied thus far. The authors suspect
this has to do with the under-determined nature of the DSE. In general, we also find that for each model studied the bootstrap method converges much faster for either $g$ greater than or less than the critical point, and much slower on the other side. This can also be seen in bootstrapped matrix models.

Example code for the generation of invariants, the DSE, and positivity constraints from the sub-matrix of $\mathcal{M}$ of a fixed size, along with implementation in \textit{Matlab} can be found on the arXiv page of this article. In subsequent plots we refer to the positivity constraints by specifying the size of the submatrix used to derive the constraints.

\section{Bootstrapped results}\label{sec:results}
Here we bootstrap a quartic and two hexic tensor models. To make notation more compact,
we no longer write out the traces, but only the graphs, e.g.  $\vone$ instead of $\operatorname{Tr}_{\vone}(T,\bar T)$.

\subsection{The quartic model}\label{sec:1of4}

For real coupling $g$, the action of the quartic model reads
\[
S(T,\bar T)=
N^{s_0}
\littlemelon
+
\frac{N^{s_1}}{2}  g\big(
\vone + \vtwo+ \vthree
\big)\,.
\]
where $s_0$ and $s_1$ are scaling factors to determine (to match
the notation in Eq. \eqref{expectation_vals}, we stress that $2C_1=\vone, 2C_2= \vtwo, 2C_3= \vthree$).
Let us define the moments $m_*$ through
\begin{subequations}%
\[
  m_2& := \ErW{ \tfrac1N \littlemelon}  \\[3pt]
 m_{2|2}&:= \ErW{\tfrac 1{N^2}\littlemelon\dot\cup \littlemelon} \label{m22}\\[3pt]
m_4&: = \ErW{\tfrac1N\vone }= \ErW{\tfrac1N\vtwo} =  \ErW{\tfrac1N\vthree }\\[3pt]
m_6& :=  \ErW{\tfrac1N\Qone }= \ErW{\tfrac1N\Qtwo} =  \ErW{\tfrac1N\Qthree }\\[3pt]
m_{6,\text{d}}& :=\mathbb E_g [\tfrac1N\Eone ] =\mathbb E_g [\tfrac1N\Etwo] =\mathbb E_g [\tfrac1N \Ethree]\\
m_{6,\text{h}}&: =\mathbb E_g[  \kthree] \label{mkthreethree}
\]
Notice that some of these equalities are not definitions; they follow from
 the fact that all quartic interaction vertices feature
 the same coupling constant. Whenever possible, we thus simplify
 omitting labels when permuting  different colourings
 do not change the expectation value:
 \[
m_{8}&: =\mathbb E_g [\tfrac1N\Aneutr ] 
&&& m_{8,\text{s}}&: =\mathbb E_g [\tfrac1N\Sneutr]\\ 
 m_{8,\text{m}}&: =\mathbb E_g [\tfrac1N\!\!\Mokka]  &&& 
m_{8,\text{x}}&: =\mathbb E_g [ \!\X] \label{meightX}\\
m_{8,\text{r}}&: =\mathbb E_g [\tfrac1N\!\Rneutr]  &&&
m_{8,\text{t}}&: =\mathbb E_g [\tfrac1N\!\Triple]
\]
\end{subequations}%

Inserting the quadratic invariant into Eq. \eqref{DSE} one obtains%
\begin{subequations}\[
N^{s_0+1 } m_2 +3g N^{s_1+1} m_4 & = N^3,
\]
and concludes that $s_0=s_1=2$ in order for a non-trivial large-$N$ limit to exist.
Then, the DSE's  Eq. \eqref{DSE} for the next five invariants in \eqref{list}
(that do not differ by colouration) read
\[
m_2 +3g  m_4&=1 \\
 m_4 +g  (m_6+2m_{6,\text{d}} ) &=(1+\tfrac 1N) m_2 \\
m_{2|2}  + 3 g  m_{4|2} & = (1+\tfrac1{N^3})m_2
  \\
 m_6 +  g (m_8+2m_{8,\text{m}} ) &=    (1+\tfrac 1N) m_4  + \tfrac 1 {N}m_{2|2}  \\
3   m_{6,\text{d}} +g  (4 m_{8,\text{m}}+2m_{8,\text{r}} +2m_{8,\text{t}}
+ m_{8,\text{s}}  ) &=
 2  \left(1 + \frac{2}{N} + \frac{1}{N^2}\right) m_4  +   m_{2|2}.
\]\end{subequations}%

\color{black}
For the following argumentation, DSE for the rest of the
invariants in \eqref{list} is also considered, but for the sake of
exposition they are presented in the Appendix.

In \cite{nguyen2015analysis}, the authors computed observables in the
large $N$ limit, as well as some corrections, of the $T^{4}_{d}$-model
by using intermediate field theory to represent such tensor
observables as the observables of a matrix integral and then applying
a saddle point analysis. The critical phenomena of said model were
later studied in \cite{delepouve2015phase,benedetti2015symmetry} and
further corrections to the large $N$ limit were computed using blobbed
topological recursion \cite{bonzom2018blobbed}. In particular, they
found two possible solutions for $m_{2}$. In the notation of this
paper they are
    \begin{equation*}
        m_{2} = \begin{cases}
      \frac{-1-\sqrt{1+12 g}}{6g} & g\leq -\frac{1}{12} \\
      \frac{-1+\sqrt{1+12 g}}{6g} & g\geq -\frac{1}{12}.
   \end{cases}
    \end{equation*}
    The solution with a minus sign is the melonic solution, and it enumerates melonic graphs, while the solution with the positive sign is known as the instanton solution. 

\begin{figure}\centering
\subfloat[]{\label{a}\includegraphics[width=.45\linewidth]{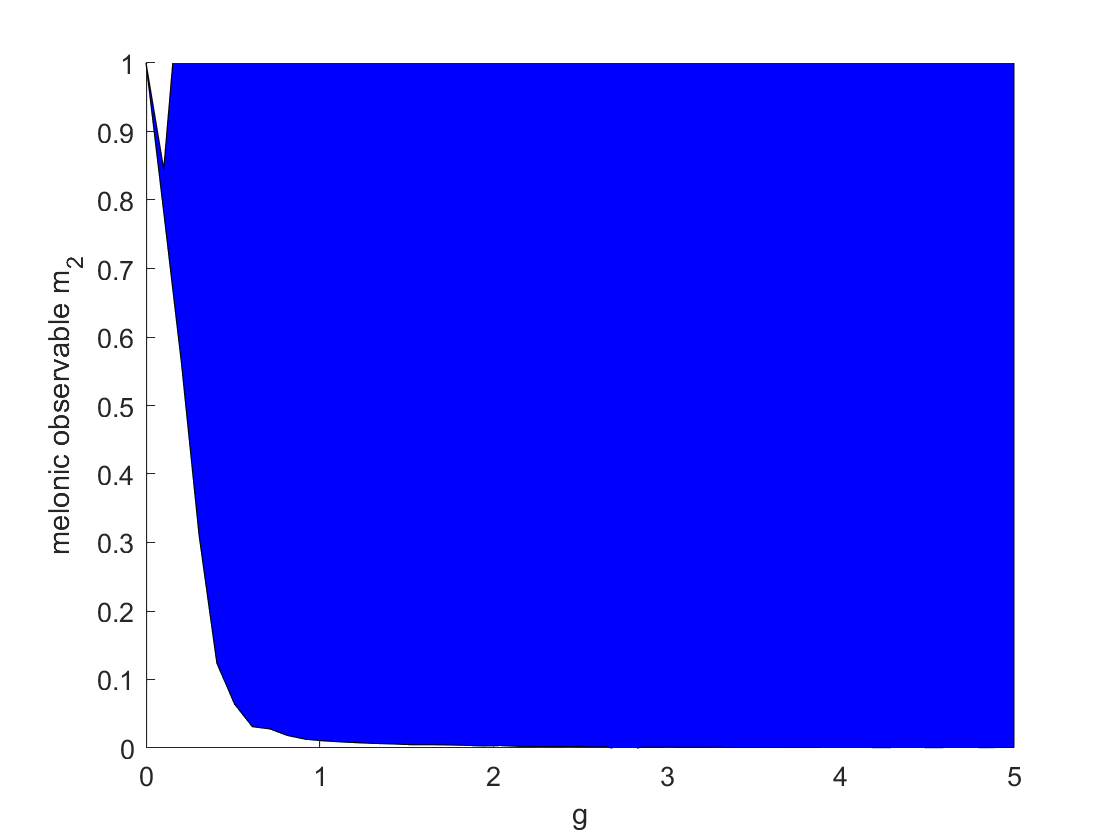}}\hfill
\subfloat[]{\label{b}\includegraphics[width=.45\linewidth]{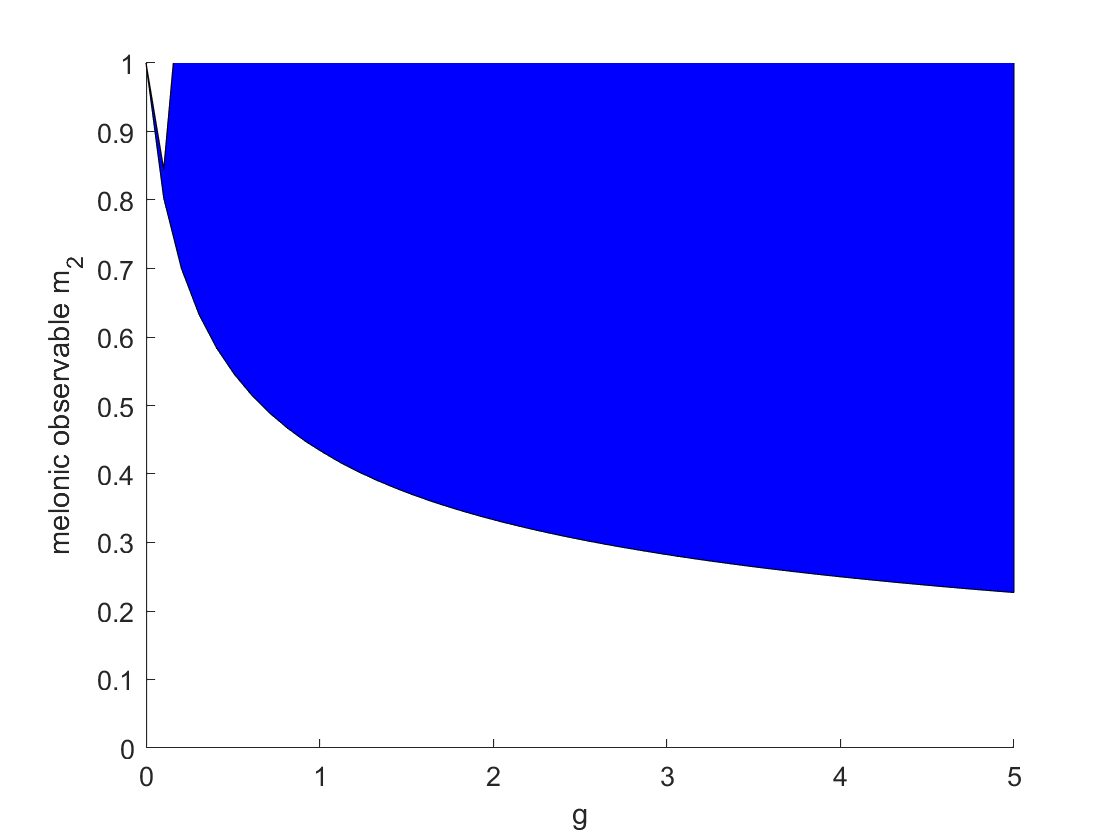}}\par 
\subfloat[]{\label{c}\includegraphics[width=.45\linewidth]{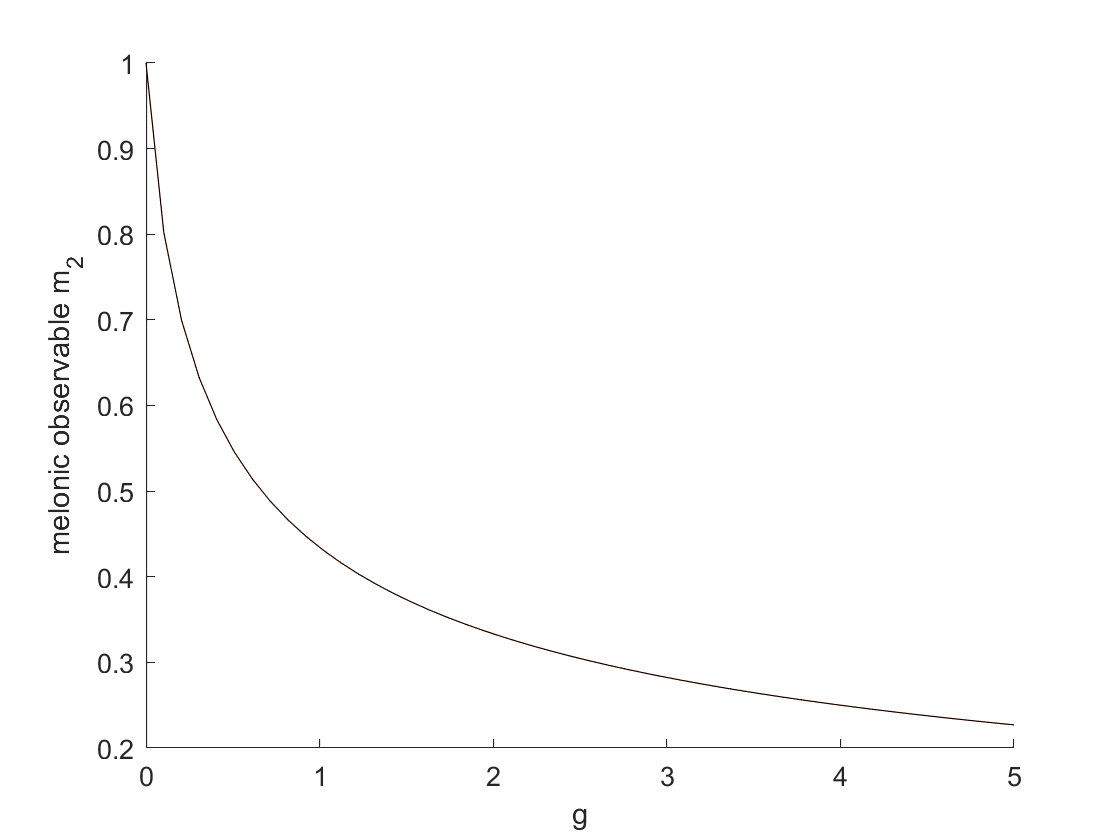}}

\caption{Bootstrapped solution space of $m_{2}$ for the quartic melonic rank three tensor model. Sub-figure (a), (b), and (c) use the constraints that $\mathcal{W}_{[2]}\geq 0$, $\mathcal{W}_{[4]}\geq 0$, and $\mathcal{W}_{[4]},\mathcal{M}_{[2]}\geq 0$, respectively with twenty-three DSE generated from graphs of increasing number of vertices starting at the melon. See the Table \ref{tab:Graphs} for a detailed list.
  \label{fig:quartic_m2}}
\end{figure}

\begin{figure}[H]\centering
\subfloat[]{\label{a}\includegraphics[width=.45\linewidth]{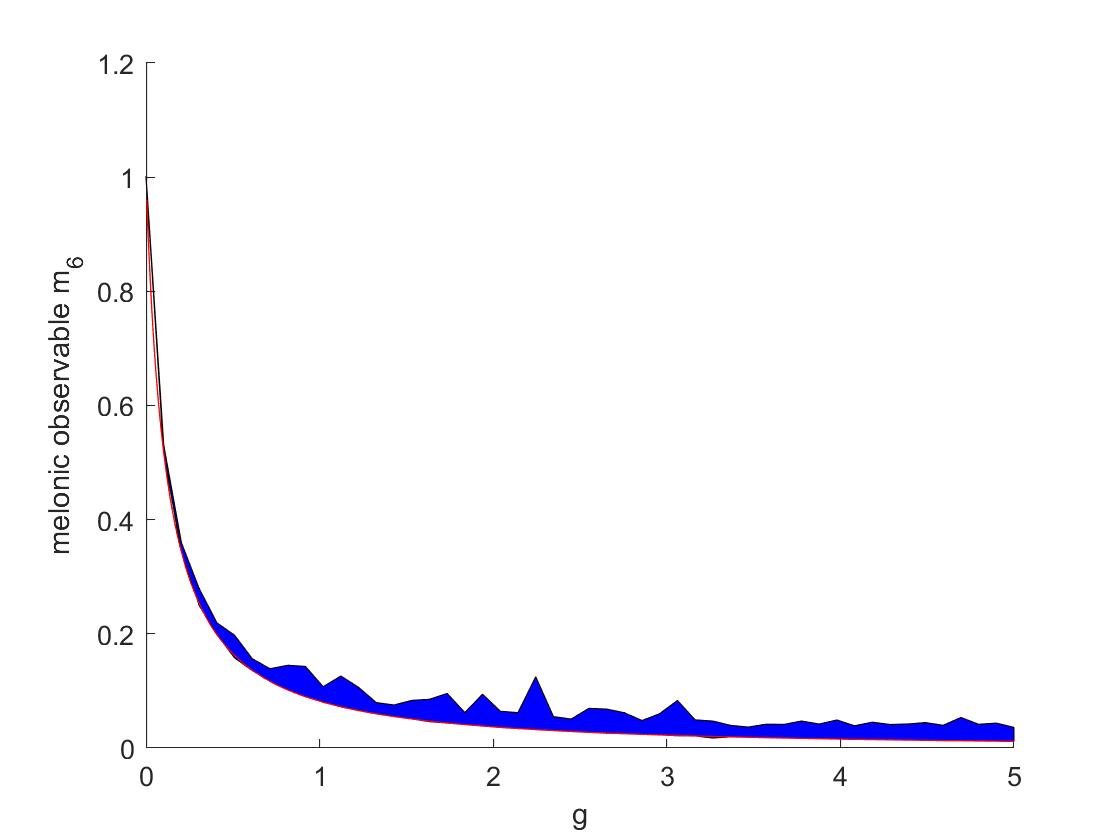}}\hfill
\subfloat[]{\label{b}\includegraphics[width=.45\linewidth]{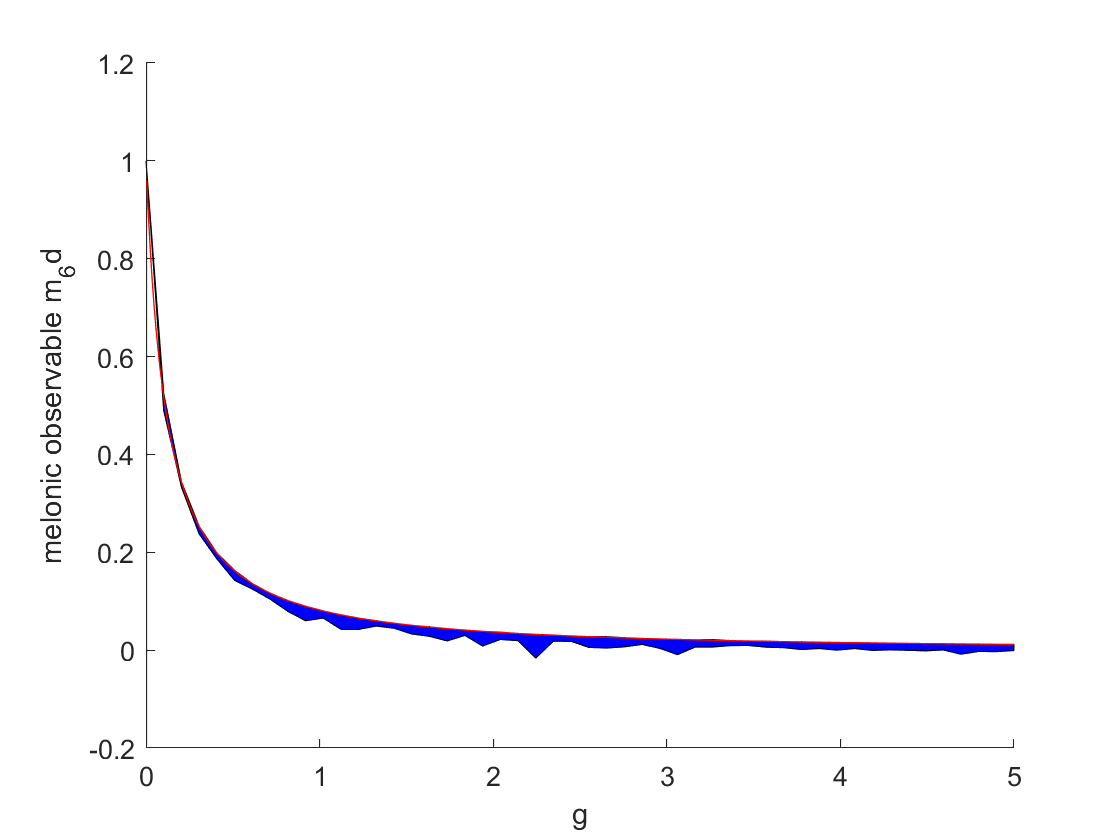}}\par

\caption{Bootstrapped solution space of $m_{6}$ and $m_{6d}$ for the quartic melonic rank three tensor model. Both figures use the constraints $\mathcal{W}_{[8]},\mathcal{M}_{[5]}\geq 0$. The red curve is the known analytic solution.
  \label{fig:quartic_m6}}
\end{figure}

Suppose we wish to obtain bootstrap estimates for $m_{2}$ of this model. Insisting, that principal minor $\mathcal{M}_{[2]}\geq 0$, allows us to obtain the strictest possible upper bound on the solution to $m_2$ in the large $N$ limit i.e. for $g\geq 0$
\begin{equation*}
    m_{2} \leq  \frac{-1+\sqrt{1+12 g}}{6g}.
\end{equation*}
The lower bounds on $m_{2}$, can then be derived from the principal minors of $\mathcal{W}$. In Figure \ref{fig:quartic_m2} we can see that after applying the constraints that $\mathcal{W}_{\{1,2,3,4\}},\mathcal{M}_{\{1,2\}}\geq 0$, the solver is able to find a curve incredibility close to the exact solution. Note that such a precise estimate for $m_{2}$, gives an equally precise estimate for $m_{4}$ from the first DSE. One can apply the same approach for other moments such as $m_{6}$ and $m_{6,d}$, but with additional constraints. See Figures \ref{fig:quartic_m6}. 

\subsection{The hexic cyclic bubble model}\label{sec:2of4}
Consider the hexic model with interaction
\[
S(T,\bar T)=
N^{2} \bigg[
\littlemelon
+
\frac{g}{3}  \big(
\Qone + \Qtwo+ \Qthree
\big) \bigg]
\,.
\]
The first few DSE are \vspace{-4ex}
\begin{subequations}\[
m_2 +3g  m_6 &=1 \\
 m_4 +g  (m_8+2m_{8,\text{m}} ) &=(1+\tfrac 1N) m_2 \\
m_{2|2}  + 3 g  m_{6|2} & = (1+\tfrac1{N^3})m_2 \\
 m_6 +  g (m_{10}+2 m_{10,\text{y}} ) &=    (1+\tfrac 1N) m_4  + \tfrac 1 {N}m_{2|2}  \\
3   m_{6,\text{d}} +g  (4 m_{10,\text{t}}+2m_{10,\text{d}} +2m_{10,\text{e}}
+ m_{10,\text{s}}  ) &=
 2  \left(1 + \frac{2}{N} + \frac{1}{N^2}\right) m_4  +   m_{2|2},
\]\end{subequations}
where \vspace{-4ex}
\[ \notag
m_{10} & = \tfrac1N \mathbb E_g\big[ \tenQ \big]
&&& m_{10,\text t} &= \tfrac1N \mathbb E_g\big[ \tent \big]
\\
m_{10,\text  s} &= \tfrac1N \mathbb E_g\big[ \tens \big]
&&&\notag
m_{10,\text y} & = \tfrac1N \mathbb E_g\big[ \teny \big]
\\
m_{10,\text e} & = \tfrac1N \mathbb E_g\big[ \tene \big]
&&&\notag
m_{10,\text d} & = \tfrac1N \mathbb E_g\big[ \tend \big].
\]

\allowdisplaybreaks[4]

In \cite{lionni2019multi}, it was shown that $m_{2}$ satisfies the equation polynomial equation $x + 3 t_6 x^3 = 1$ in the
limit. Thus, we have that
\begin{equation*}
m_{2} =\frac{1}{3} \left(\frac{\sqrt[3]{9 t_{6}^2+\sqrt{81
   t_{6}^4+4 t_{6}^3}}}{\sqrt[3]{2}
   t_{6}}-\frac{\sqrt[3]{2}}{\sqrt[3]{9 t_{6}^2+\sqrt{81
   t_{6}^4+4 t_{6}^3}}}\right).
\end{equation*}
Using the first DSE, one can then obtain a formula for $m_{6}$. 
In Figure  \label{fig:hexic_m6}, we see excellent agreement
between the analytic solution for $m_{6}$ and the bootstrapped
solution.  
 
It is worth, as an aside, to use this result to write out the free
energy. Combining this relation with the DSE coming from the pillow graphs and the formula for $m_{2}$, implies for the hexic cyclic model that $m_{6} =
m_{2}^3$ and for the hexic pillow model that $m_{6,d} = m_{2}^3$. Note
that
$$\frac{1}{\alpha}m_{6} = \frac{d}{d g}\log Z,$$
 where $\alpha =3$ for the hexic cyclic model and $\alpha=2$ for the hexic pillow model. Hence, the free energy can be written explicitly as
\begin{align*}
  \lim_{N \rightarrow \infty}\frac{1}{N^3}\log Z(N,t_{6})&=\int^{t_{6}} \frac{1-m_{2}}{3 t} dt 
  \\
  &=\int^{t_{6}}\frac{1}{{18 t^2}}\left[6 t +\frac{2 \sqrt[3]{2} t}{\sqrt[3]{\sqrt{t^3
   (81 t+4)}+9 t^2}}-2^{2/3}
   \sqrt[3]{\sqrt{t^3 (81 t+4)}+9 t^2}\right]dt,
\end{align*}
where $\int^{g}$ denotes the antiderivative.

 Unlike with the quartic model, here we find that we are able to find excellent agreement with the known analytic solution only using constraints from the $\mathcal{W}$ matrix.  See Figure \ref{fig:hexic_m6}.

\begin{figure}[H]\centering
\subfloat[]{\label{a}\includegraphics[width=.45\linewidth]{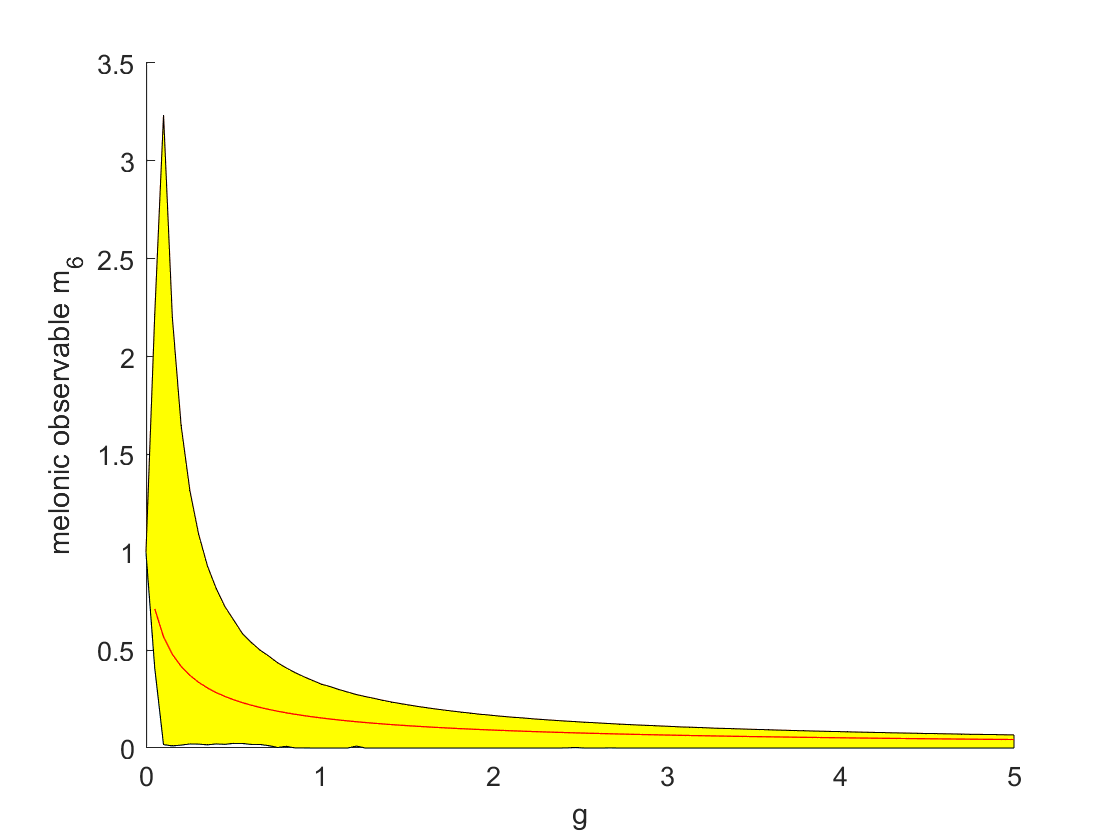}}\hfill
\subfloat[]{\label{b}\includegraphics[width=.45\linewidth]{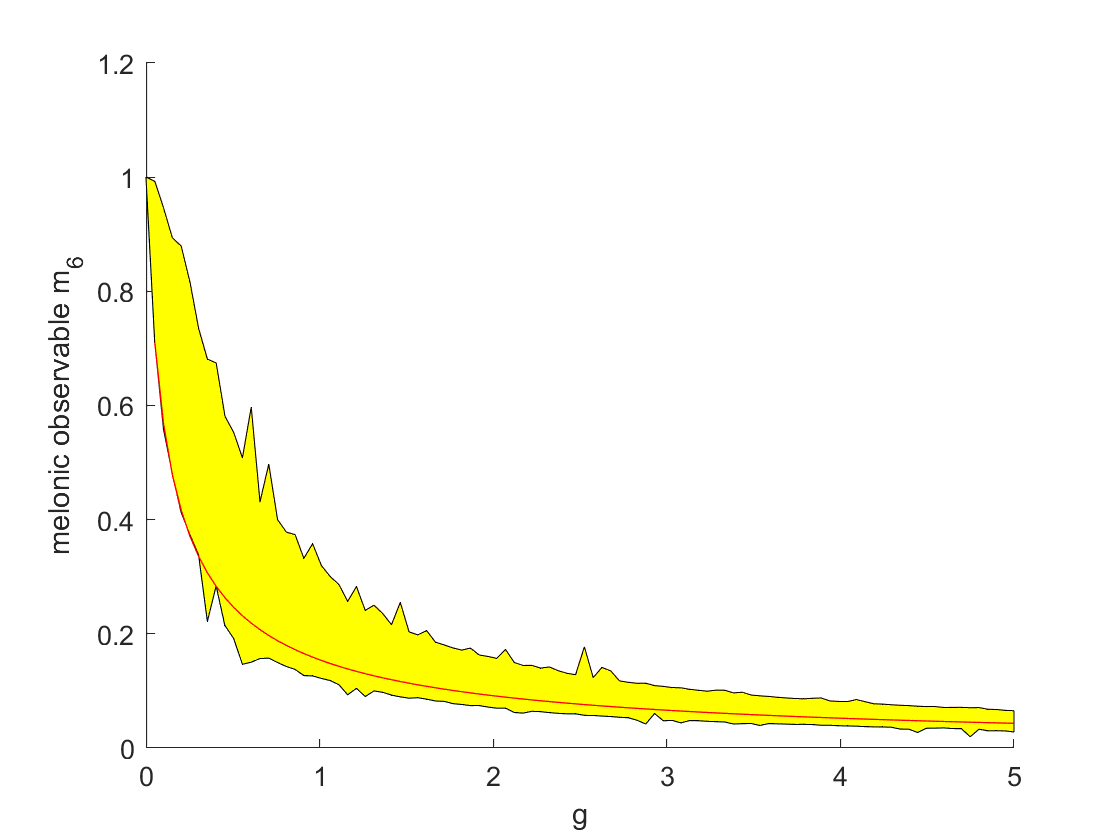}}\par 
\subfloat[]{\label{c}\includegraphics[width=.45\linewidth]{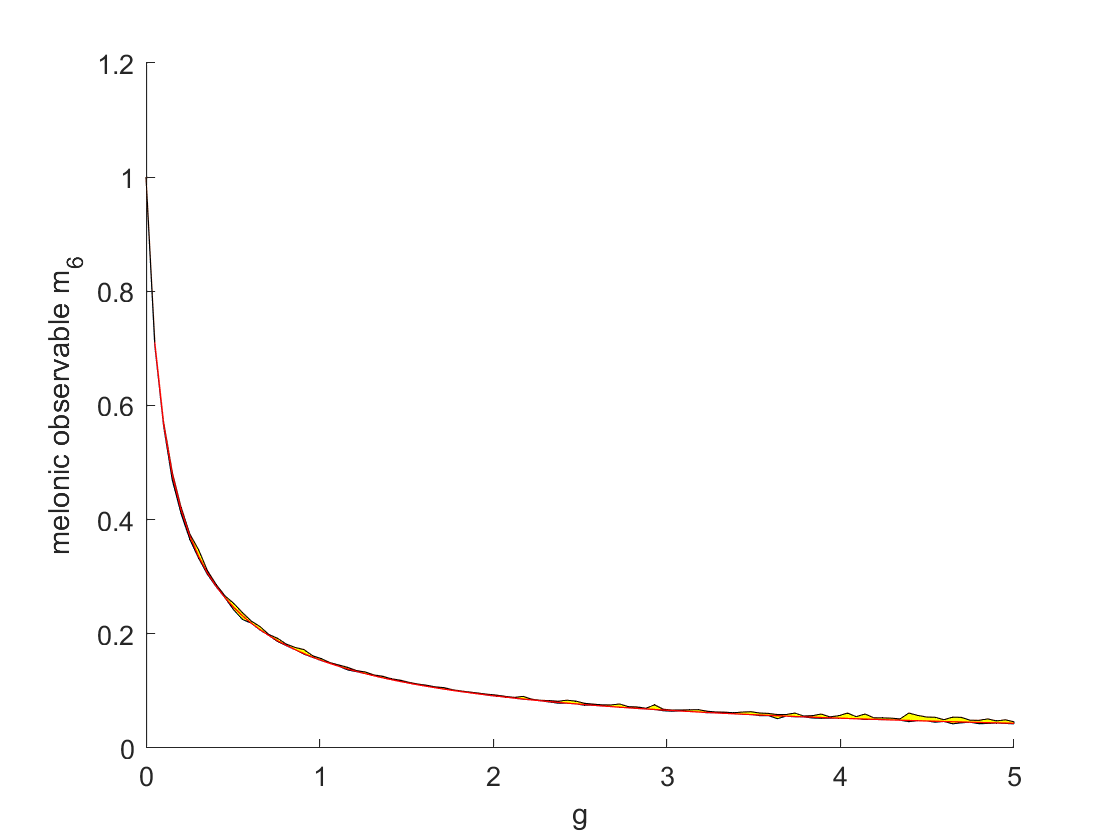}}

\caption{Bootstrapped solution space of $m_{6}$ for the heixc cyclic bubble rank three tensor model. Sub-figure (a), (b), and (c) use the constraints that $\mathcal{W}_{[4]}\geq 0$,$\mathcal{W}_{[6]}\geq 0$, and $\mathcal{W}_{[8]}\geq 0$, respectively, with twenty DSE generated from graphs of increasing number of vertices starting at the melon. See Table \ref{tab:Graphs} for a detailed list.
  \label{fig:hexic_m6}}
\end{figure}

\section{A conjecture for the cumulants of the quartic model}\label{sec:conjecture}

Consider again the quartic model
\[
S_g (T,\bar T)=
N^{2} \bigg[
\littlemelon
+ \frac g 2 \big(
\vone + \vtwo+ \vthree
\big) \bigg ] \,.
\]
Given a 3-coloured graph $B$, one defines
the formal connected expectation value $ \mathbb E_g\conn [B] $ by
\[
\mathbb E_g\conn [B]  &: =  \sum_{n\geq 0} \Big(\frac {-g} 2 \Big)^n
 \int\conn \frac {N^{2n} } {n!}
 B \cdot \big  (\vone+\vtwo + \vthree  \big)^n \dif \mu_N (T) \notag  \\
  & =   \sum_{ \substack { n, k_1,k_2,k_3\geq 0 \\ k_1+k_2+k_3=n }}\Big(-\frac {gN^2} 2 \Big)^n
   \int\conn  B \cdot
 \frac{  (\vone)^{k_1} \cdot (\vtwo)^{k_2} \cdot (\vthree)^{k_3} }{k_1!  \cdot k_2! \cdot k_3!} \dif \mu_N (T, \bar T),
 \label{expansionEg} \]
 where $\dif \mu_N(T, \bar T)$ is the Gau\ss ian
 tensor measure $\exp(-N^2\littlemelon) \dif T\wedge \dif \bar T$
 over $\C^{N^3}$ as defined in Eq. \eqref{measureT}. Here, `conn.' means that at the time of
 evaluating the integral in terms of Feynman graphs, one keeps
 only the contributions by the connected ones.
\\

\noindent \textbf{Conjecture I.} \textit{Let $B$ be a connected 3-coloured graph. Define  $
I_{i_1,\ldots,i_q; N} [B]$ $(q\in \Z_{>0})$ by }
\[
I_{i_1,\ldots,i_q; N}[B]
= N^{2q-1}   \int\conn   \big(  \maketrace B \cdot  \vione \cdots \viq  \big) \dif \mu_N(T, \bar T) \qquad i_1,\ldots,i_q \in \{1,2,3\}.
\]
\textit{Then  $\mathrm{L.O.}[
I_{i_1,\ldots,i_q; N}]$, the leading order in $N$ of $I_{i_1,\ldots,i_q; N}$,
 depends only on the set $\{i_1,i_2,\ldots,i_q\}$ through its
cardinality $q$}.
\\

We comment on the meaning of this statement.
\begin{itemize}
\item The indices of the tensors being three is essential.

\item The previous  statement should not be confused with the following one:
If $\{\tau (i_1),\tau (i_2) \ldots,\tau (i_q) \} $ and $i_1,i_2,\ldots, i_q$ are
equal as sets for some
permutation  $\tau \in \Sym(3)$, then
\[
I_{\tau (i_1),\tau (i_2) \ldots,\tau (i_q); N}[B]=
I_{i_1,\ldots,i_q; N}[B],
\]
which is true, but also trivial to prove (say by observing that
the $c$-coloured loops on the LHS correspond to $\tau(c)$-coloured
loops in the RHS); see also the next point.

\item In fact, even for a graph $B$ that has no preferred colour---i.e. which is preserved by the action of $\Sym(3)$, like the
 two-vertex melon---the conjecture is not an empty statement.
We illustrate it now by explicit computation of the two point function double series in \texttt{feyntensor} \cite{feyntensor}:
\end{itemize}
\allowdisplaybreaks[4] {
\footnotesize
\[  \nonumber
 \qquad \mathbb E_g\conn [ \tfrac 1N \littlemelon ]
& = 1 - {3g}  \Big( 1+ \frac1N \Big)\\ \notag
&
+ {3g^2} \bigg[ \frac 1{2^2 \cdot 2!}  \Big(  \mathbf{16} + \frac{40}{N} + \frac{16}{N^{2}} + \frac{8}{N^{4}} \Big)+ \frac1{2^2}    \Big(\mathbf{16} + \frac{32}{N} + \frac{24}{N^{2}} + \frac{8}{N^{3}}   \Big) \bigg]
\\ \notag
& -  {g^3}\bigg[ \frac{3 }{2^3 \cdot 3! } \Big( \mathbf{240}+
\frac{1056}{N} + \frac{1056}{N^{2}} + \frac{240}{N^{3}} + \frac{480}{N^{4}} + \frac{480}{N^{5}}  \nonumber \\ \notag
&  +  \frac{6  }{2^3 \cdot 2!\cdot  1!  }  \Big( \mathbf{240}+ \frac{816}{N} + \frac{1056}{N^{2}} + \frac{720}{N^{3}} + \frac{480}{N^{4}} + \frac{240}{N^{5}}   \Big) \\ \notag
&  +  \Big( \mathbf{240}+ \frac{720}{N} + \frac{1008}{N^{2}} + \frac{1056}{N^{3}} + \frac{432}{N^{4}} + \frac{96}{N^{6}}
\Big) \bigg] \nonumber
\\ \notag\notag
& +  {g^4}\bigg[ \frac{3}{  2 ^ 4 \cdot  4! } \Big(\mathbf{5376} + \frac{35712}{N} + \frac{62976}{N^{2}} + \frac{35712}{N^{3}} + \frac{32256}{N^{4}} + \frac{64128}{N^{5}} + \frac{26880}{N^{6}} + \frac{8064}{N^{8}} \Big) \\ \notag  \nonumber &
+\frac{6}{2 ^ 4  \cdot 3! \cdot  1! }  \Big( \mathbf{5376} +\frac{27648}{N} + \frac{52608}{N^{2}} + \frac{54144}{N^{3}} + \frac{50688}{N^{4}} + \frac{45696}{N^{5}} + \frac{26880}{N^{6}} + \frac{8064}{N^{7}} \Big) \\ \notag   \nonumber&
+\frac{3}{2 ^ 4  \cdot 2! \cdot  2! }   \Big( \mathbf{5376}+ \frac{25088}{N} + \frac{50432}{N^{2}} + \frac{56960}{N^{3}} + \frac{53760}{N^{4}} + \frac{53504}{N^{5}} + \frac{21504}{N^{6}} + \frac{4480}{N^{8}}  \Big) \\ \notag  \nonumber &
+\frac{3}{2 ^ 4  \cdot 2! \cdot  2! }    \Big( \mathbf{5376} +
\frac{23296}{N} + \frac{46336}{N^{2}} + \frac{64512}{N^{3}} + \frac{63232}{N^{4}} + \frac{38016}{N^{5}} + \frac{20608}{N^{6}} + \frac{9728}{N^{7}}
\Big) \bigg]  + O(g^5)\,. \notag
\]
}

\normalsize\noindent
\begin{itemize}
\item[]  
Observe that the bold terms are of the form L.O.$( I_{i_1,\ldots,i_q} [\tfrac 1N \littlemelon] )$ (for some $q$). The conjecture means that
 L.O. $I_{k_1,k_2,k_3,k_4 ; N} [\tfrac 1N \littlemelon] $ is determined by  $N^7
 \int\conn \big( \littlemelon\cdot (\vone)^4 \big)$ alone  (in view of the previous point,
 this is trivial if the $k_i$ are all equal to 2 or 3, but it is not obvious
 if at least any two indices differ),
 and there is --- to the knowledge of the authors --- no
 available proof. It is not clear either how it follows from the Feynman diagrammatics
 that the only dependence of $I_{k_1,k_2,\ldots, k_q ; N} [\bullet] $ on $k_1,\ldots, k_q$ is through $q$.

\item In the conjecture $B$ can be non-melonic and of any genus.
\end{itemize}
\noindent
For future reference, we provide the re-ordered expression in the Appendix \ref{app:expansions}, which
is easily seen to  coincide with the series $(-3)^n \textrm{Cat}_n $
generated by $-\frac{\sqrt{1+12 \, g } + 1}{6 \, g}$; this is
precisely the series expansion of the solution found in
\cite{nguyen2015analysis}, discussed in the previous section.

Let $\mathrm{Cycl}_c(p)$ denote the cyclic melonic graph built from
the two-vertex melon $\littlemelon$ by $p-1$ insertions of a
$c$-coloured dipole,  for instance, \[\mathrm{Cycl}_1(2) =\vone ,\quad
\mathrm{Cycl}_2(3) = \Qtwo ,\quad \mathrm{Cycl}_1(4) = \Aone , \quad
\mathrm{Cycl}_3(5) =\tenQ\vphantom{a}^{\!\text{\tiny 3}}, \ldots \]
\noindent \textbf{Conjecture II.} \textit{Let $M_1,\ldots,M_q$ be  melonic, connected 3-coloured graphs, $q\in \Z_{>0}$. Define  $
J_N  ( {M_1},\ldots,{M_q} )$ by}
\[
J_N ( {M_1},\ldots,{M_q})
= N^{2q-3}   \int\conn    \Tr_{ M_1}(T,\bar T) \cdot \Tr_{ M_2}(T,\bar T)
\cdots \Tr_{ M_q}(T,\bar T)  \dif \mu_N(T, \bar T).
\]
\textit{Then }
\[ \label{strong_conj} \lim_{N\to \infty} J_N  ( {M_1},\ldots,{M_q} ) =   \lim_{N\to \infty}  J_N  \big( {\mathrm{Cycl}_1(p_1)},
\mathrm{Cycl}_1(p_2) , \ldots, \mathrm{Cycl}_1(p_q)  \big)    \]
\textit{where $2p_i=\# V(M_i)$ for each $i=1\ldots, q$. The R.H.S. is independent of the colouration (hence we let all colours be equal to $1$).}
\\

Conjecture II was later proven in \cite{TwofoldUniversality} but introducing the needed framework would lead astray.  We content ourselves here by giving the evidence that supports it. \\ 

\noindent
\textit{Support of the conjectures.} Let
\[ J_{p_1,p_2,\ldots, p_q} := \lim_{N\to \infty}  J_N  \big( {\mathrm{Cycl}_1(p_1)},
\mathrm{Cycl}_1(p_2) , \ldots, \mathrm{Cycl}_1(p_q)  \big). \label{defJq}
\]
By explicit computation with \texttt{feyntensor} \cite{feyntensor} (written in \texttt{SageMath} \cite{sagemath}) we found the following values:
\begin{subequations}%
\[
J_{2,3} & =6 &&& J_{2,3,3}&= 126   &&& J_{2,2,3,3} &= 2592  \\
J_{3,3} & =9 &&& J_{3,3,3}&= 216  &&& J_{3,4} &= 12 \\
J_{2,2,3} & = 72  &&& J_{2,2,2,3}&=1344   &&& J_{4,4} &= 16 \\
&  &&& J_{2,3,4}  &= 192 && &  & \]\label{conj_support}%
\end{subequations}%
\noindent
To support the conjecture, we computed the LHS of Eq. \eqref{strong_conj}
for any set of melonic connected graphs $M_1,\ldots, M_q$ whose vertex numbers
are determined by Eqs.  \eqref{defJq} and \eqref{conj_support}. All verified the equality  \eqref{strong_conj}.
For example,  the number $J_{2,2,3,3} =2592$ was also the LHS of  \eqref{strong_conj}
with $M_1,\ldots, M_4 $ any of the listed invariants:
\begin{align*}
(M_1,M_2,M_3,M_4) \in \Big\{
(\Eone,\Qtwo,\vone,\vtwo), &&& ( \Eone,\Qtwo,\vtwo,\vtwo), &&  ( \Eone,\Qtwo,\vone,\vthree) \\ ( \Eone,\Qtwo,\vthree,\vthree), &&&
         (\Eone,\Eone,\vone,\vtwo), && ( \Eone,\Eone,\vtwo,\vtwo), \\ ( \Eone,\Eone,\vtwo,\vthree), &&& ( \Eone,\Eone,\vone,\vone), &&
         (\Eone,\Etwo,\vone,\vtwo), \\ ( \Eone,\Etwo,\vtwo,\vtwo),  &&& ( \Eone,\Etwo,\vone,\vthree), && ( \Eone,\Etwo,\vthree,\vthree), \\
         ( \Eone,\Eone,\vthree,\vthree), &&& ( \Qone,\Qone,\vtwo,\vtwo),  && ( \Qone,\Qone,\vtwo,\vthree), \\ ( \Qone,\Qone,\vtwo,\vthree),  &&&
         (\Qone,\Qtwo,\vone,\vtwo), && ( \Qone,\Qtwo,\vtwo,\vtwo), \\ ( \Qone,\Qtwo,\vone,\vthree), &&& ( \Qone,\Qtwo,\vthree,\vthree),  &&
        (\Qone,\Qone,\vone,\vtwo)  \Big\}
\end{align*}
Further, none of the integrals performed agreed beyond the leading order. $\hspace{4.5cm}\diamond$ \\

\begin{figure}[H]
\begin{minipage}{.5\textwidth}
\includegraphics[width=.95\textwidth]{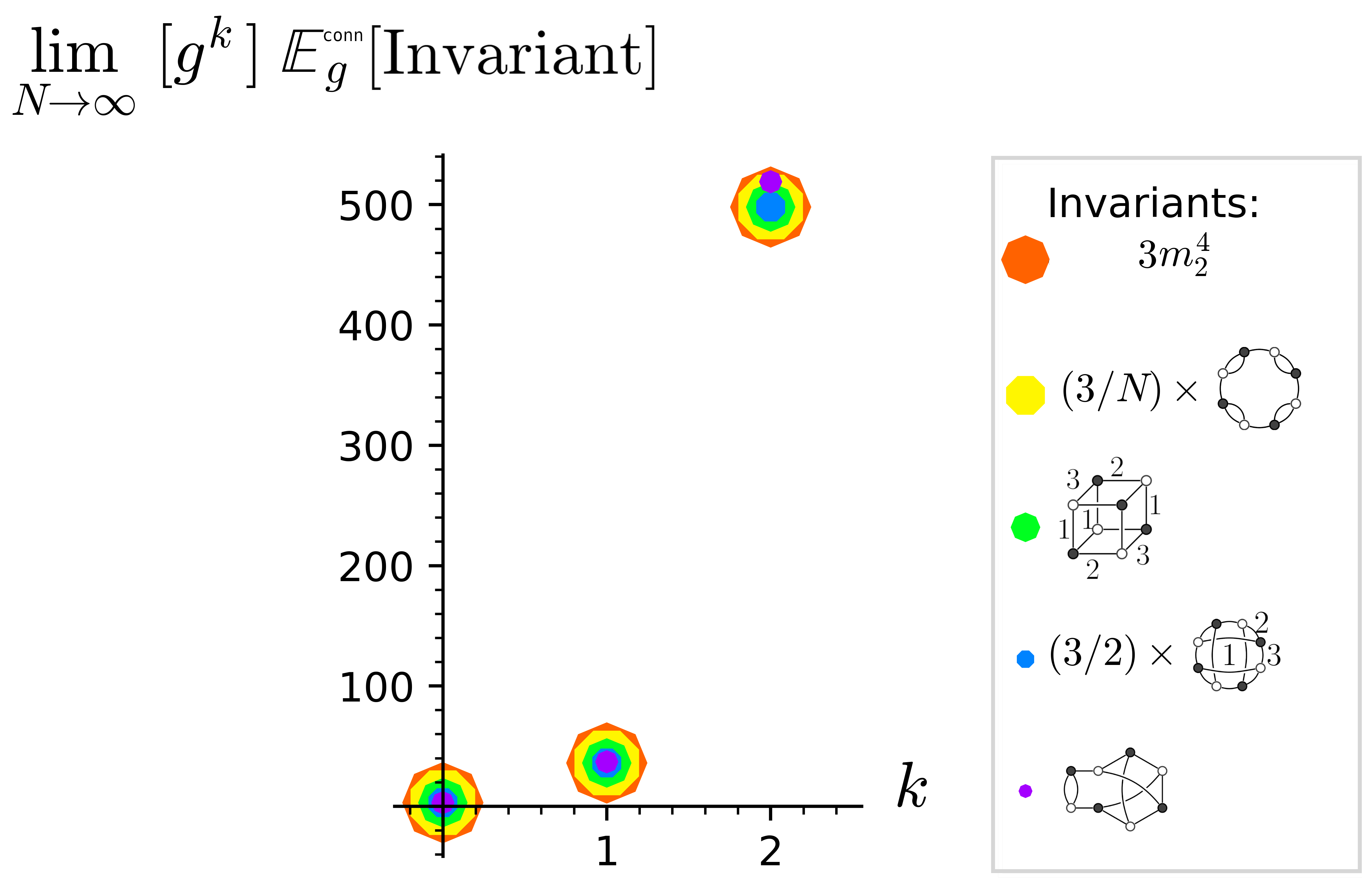}
\end{minipage}
~
\begin{minipage}{.45\textwidth}
\caption{This plot shows the $g^k$ coefficients of leading-orders of
  moments computed in the ensemble with genus-1 hexic interaction with
  coupling $g$.  On the non-factorization of tensor invariants: in
  this table on the right, $m_2$ is the two-point function.  We plot
  the exponents in $g$ of each correlator. Notice that all graphs
  except the last two are planar; the third graph is non-melonic but
  planar, see \eqref{correlators_k33_intro}.\label{fig:coeffs}}
\end{minipage}
\end{figure}

The point of the very special role of the genus, as commented on in the Introduction, can be concretely illustrated in the double series expansion of the correlators
in the hexic model 
with the potential $
S_g(T,\bar T)=
N^{2} \big(
\littlemelon
+ N \frac g3 \kthree  \big) $ whose interaction graph $\kthree $ 
(this is the genus-1 graph
presented in Figure \ref{fig:mug}; knowing that this graph is coloured allows us to drop
the colour labels).
Consider now the following three octic correlators, for which \texttt{feyntensor} yields:
\allowdisplaybreaks[0]
\begin{subequations} \label{correlators_k33_intro}
\[
\mathbb E_g \conn [\tfrac1N   \Aneutr ]&=  1 - 12 g + 166 g^{2} +  O(g^3,N^0) +  \left(6 - 96 g + 1674 g^{2}+\ldots\right) \frac1N  + O(1/N^2) \\
\mathbb E_g \conn [ \!\! \Y ] &=
2  - 24 g + 332 g^{2}  + O(g^3,N^0)  + \left(8  - 140 g + 2582 g^{2}+ \ldots\right) \frac1N   +O(1/N^2)
\\
\mathbb E_g\conn [  \Cubon] &=
3 - 36 g + 498 g^{2} + O(g^3,N^0) + \left(6 - 132 g + 2748 g^{2}+\ldots\right) \frac1N  +O(1/N^2).
\]
This expansion suggests that they have a leading order in $N$ that allows them to be a common multiple of
the two point function below, to the fourth power 
$m_2^4=  1  - 12g  +166 g^2 $, where $m_2 $ is the large-$N$ limit of
 \[  \nonumber
 \mathbb E_g\conn [ \tfrac 1N \littlemelon  ]
& =1 -
3 \, g {\left(\frac{1}{N} + 1\right) }+ g^{2} {\left(28+ \frac{72}{N} + \frac{63}{N^{2}} 
+ \frac{42}{N^{3}} + \ldots  \right)} -g^{3} {\left(351 + \ldots \right)} + \ldots \nonumber
\]
In contrast,
the following expectation value of the genus-one invariant
\[ \mathbb E_g \conn  [\!\!  \X ] &=
3 - 37 g + 519 g^{2} + O(g^3,N^0) + \left(8 - 162 g + 3259 g^{2}+\ldots\right) \frac1N +O(1/N^2) \]
will never factor, as it can be seen from the first non-constant coefficient
mismatching 36 (see Fig. \ref{fig:coeffs}).
\end{subequations}

\allowdisplaybreaks[4]

\section{Summary and future works}\label{sec:summary}
This paper introduced the bootstrapping with positivity methodology to random $U(N)$-invariant tensor models. In an analogous manner to the Hankel matrix in the Hamburger moment problem, we derived positivity constraints on a symmetric expected value of invariants. We then apply this method to the quartic and the three hexic rank three tensor models in the large $N$ limit. In all models studied, we find excellent agreement with the known analytic solutions for the expected value of the melon. Series expansions and bootstrapped results led us to conjecture that, for the quartic model in the large $N$ limit, the expectation value of an invariant only depends on the number of vertices of the corresponding 3-coloured graph. Using the DSE, we then conjecture a general formula for the expected value of all invariants of the quartic tensor model in the large $N$ limit.

With this new methodology the door to studying any tensor model of any rank is now open, given sufficient computing resources. As mentioned in the introduction, this methodology can easily be adapted to other types of invariance. As with matrix models, bootstrapping can also lead to the discovery of critical points and exponents. This may lead to connections to new and existing continuum theories. In future works, in addition to the above-mentioned directions, the authors will aim to prove our conjecture and perhaps find analogous conjectures for other tensor models.

We note that it is straightforward for the proposed methodology for $U(N)$ invariant tensor models in this paper to be changed to tensor models with $O(N)$ invariance \cite{carrozza2016n,bonzom2022double,krajewski2023double},
as well as $O(-N)$ invariant ones, which is equivalent to $\textrm{Sp}(N)$-invariance  \cite{SymplecticTensors}. The orthogonal case is  also physically interesting  in view of some $O(N)$-invariant tensor integrals modelling the large-$N$ SYK-model \cite{WittenSYK}, chiefly through the Carrozza-Tanas\u a--Klebanov-Tarnopolsky model \cite{carrozza2016n,KlebanovTarnopolsky}. The Dyson-Schwinger equations exist for such models and the positivity conditions follow in an analogous manner. For more models whose
symmetry is broken at the level of the quadratic term by a  `Laplacian', in the sense of generalizing the Grosse-Wulkenhaar matrix model \cite{gw12} or  Kontsevich's to tensors, loop equations \cite{fullward,SDE,SDE_disconnected} also exist,
but are more complicated.

\section*{Acknowledgements}
N. Pagliaroli would like to thank Samuel Laliberte for helpful advice and discussions.  N. Pagliaroli acknowledges the support of the Natural Sciences and Engineering Research Council of Canada (NSERC). We acknowledge the usefulness of an initial code by Niels Gehrig, 
which was here substantially extended. C. Perez-Sanchez's
work was mainly supported by the Deutsche Forschungsgemeinschaft (DFG, German Research Foundation) under Germany’s Excellence Strategy EXC 2181/1 - 390900948 (the Heidelberg STRUCTURES
Excellence Cluster).

\section*{Data availability}
The code used to reproduce the figures and computations seen throughout the article will be shared upon request. General code to implement the bootstrapping technique used here can be found on the arXiv page for this paper. Also the code \texttt{feyntensor} to perform tensor integrals is available at \cite{feyntensor}. 


\appendix

\section{Scaling factors algorithm} \label{app:scaling}

\allowdisplaybreaks[2]

Let us abbreviate each invariant $\maketrace{B}$ as its associated coloured graph $B$ in this Appendix. Its aim is to find the missing variable $t(B)$ in the
scaling $ {N^{t(B)- \# \pi_0 (B) } } \ErW{B}$ that yields a large-$N$
limit of order $N^0$ for this expression, which simultaneously is
non-trivial and has infinitely many contributions as a series in
the coupling $g$. Recall that $\pi_0 (B) $ are the connected
components of $B$.

For a connected invariant graph $B$, we recursively define
$t(B)$ by the following `pruning' process. A $d$-dipole of $B$ is by
definition a subgraph of $B$ with two vertices that are connected by
exactly $ d$ edges.  Let us denote by $d_{\max}(B) $ the maximal
dipole number found at $B$. For instance
\[d_{\max}(\!\!\X )=2,  d_{\max}(\kthree)=1
\text { and, } d_{\max} (\vone)=2. \notag \]

By pruning out of $B$ a $d$-dipole with vertices $\{v,w\}$ we mean to
remove it and to weld colourwise the remaining $D-d$ edges in $B$ that
were connected to $v$ or $w$.  The algorithm to find $t(B)$ reads then
as follows:
\begin{enumerate}
\item $t(B)=0$ if $B=\littlemelon$.
 \item Else, prune $B$ at all dipoles  $\{u,v\}$, which in each case yields
 $(B \setminus \{u,v\})\big|_{\text{\tiny welded}}$, and take the maximum:
 \[ \notag
 t(B) & = \max_{ \substack{ u \in  \Vb(B)  \\ v \in  \Vw(B) }  }   \big[ D-1-  \# E(u,v)  \\
 &\qquad + \sum_{ K \in \pi_0   [ B \setminus \{u,v\}|_{\text{\tiny welded}} \normalsize] } t(K)  \big] \,.\label{step}
 \]

\end{enumerate}

Observe first that $t(B)=0$, whenever $B$ is any melonic graph. Since $B$ is then indexed by successive insertions of $(D-1)$-dipoles into
the edges of $\littlemelon$.  If we prune backwards that sequence of
insertions, we find at any step (say the $n$-th) dipoles maximizing
the number of edges, $d_{\max}(B\hp n ) = D-1$, provided $2n< \#V(B)$
(else $B\hp n $ is the melon or empty).  After each such step the
obtained graph $B'$ is connected and a melonic insertion has been
removed. Further, \eqref{step} implies that the $t$-number is
conserved, until one reaches the two-point graph, whence $t(B)=t[B
  \hp1 ]=t[ B \hp 2]=\cdots =t(\littlemelon) $, which was initialised
to be 0.



For models of tensors with $D$ indices, the action is given by
\[
S= N^{D-1} \Big(T\cdot\bar T  + \sum_j  g_j N^{t(C_j)} C_j \Big).
\]
We suppose that the  invariants $C_j$  all have connected graphs. Multiplying  the DSE  by $N^{t(B)-\#\pi_0(B)-(D-1)}$, one obtains  ($s_0=D-1$
and $s_1=D-1 + t(C_j)$ there)
\begin{align} \frac{\#V(B)}{2}
 N^{ t(B)-
\#\pi_0 (B) }
\ErW{B } + \sum\limits_{j=1}^n
 g_{j} & N^{t(B) -\#\pi_0 (B) +
 t(C_j)  }\sum\limits_{\substack{ w \in  \Vb(C_j)  \\ v \in  \Vw(B)  } }
 \ErW{ B   \tensor[_v]{\star}{_{w}}   C_j } \nonumber \\
& =    \sum\limits_{\substack{ u \in  \Vb(B)  \\ v \in  \Vw(B)  }}
 N^{\# E  ( u, v) -(D-1) + t(B) -\#\pi_0 (B) }
 \ErW{ (B \setminus \{u,v\})\big|_{\text{\tiny welded}}}\,. \label{DSEbis}
\end{align}
\textit{If} each $C_j$ is melonic, then $t(C_j)=0$ by the above argument.
Further, since  $ 4\leq \# V(C_j) =:2p_j $ for each $j$, for arbitrary $w$ and $v$
the graph $B   \tensor[_v]{\star}{_{w}}   C_j $ has kept $p_j-1$ black vertices from $C_j$.
At least one of these lies on a dipole $\{u_i,v_i\}$,
and since removing it preserves  connectedness and the $t$-number, one has (repeating the argument)
\[
t(B   \tensor[_v]{\star}{_{w}}   C_j ) =
t(B   \tensor[_v]{\star}{_{w}}   C_j \setminus \{u_i,v_i\}|_{\text{\tiny welded}} )=
\cdots =t(B ).  \label{tpreservedbystar}
\]
Observe also that
the connected components are unchanged: $ \#\pi_0 (B   \tensor[_v]{\star}{_{w}}   C_j )   =
\#\pi_0 (B)  = \# \pi_0 [  (B \setminus \{u,v\})\big|_{\text{\tiny welded}} ]$.
Therefore, by Eqs.
\eqref{DSEbis} and \eqref{tpreservedbystar},
it follows that the leading orders  of the three terms in the loop equations coincide:
 \begin{subequations}%
\[
&\phantom{abc}  \text{L.O.}
\bigg\{ N^{ t(B)-
\#\pi_0 (B) }
\ErW{B }   \bigg\} \\[1ex] &\sim
\text{L.O.}
\bigg\{  N^{t( B   \tensor[_v]{\star}{_{w}}   C_j) -\#\pi_0 ( B   \tensor[_v]{\star}{_{w}}   C_j) }\sum\limits_{\substack{ w \in  \Vb(C_j)  \\ v \in  \Vw(B)  } }
 \ErW{ B   \tensor[_v]{\star}{_{w}}   C_j }   \bigg\}
\\  &\sim
\text{L.O.}
\bigg\{   N^{ t [ (B \setminus \{\tilde u,\tilde v\})|_{\text{\tiny welded}}    ] -\#\pi_0 [
 (B \setminus \{\tilde u,\tilde v\}) |_{\text{\tiny welded}} ] }
\bigg\},
\]
\end{subequations}
where $ \tilde u  \in  \Vb (B) ,\tilde v\in \Vw(B)$ are the vertices at which
\eqref{step} attains its maximum, and
with
\[ t [ (B \setminus \{\tilde u,\tilde v\})|_{\text{\tiny welded}}    ]  =  \sum _{K\in \pi_0 [
 (B \setminus \{\tilde u,\tilde v\})|_{\text{\tiny welded}}  ] } {   t(K) }  .\]

\section{DSE for the quartic}\label{app:SDElist}
Here we include ten DSE for the quartic rank three tensor model, whose input polynomials are of increasing size. These were generated with the Python script \texttt{DSE\_generator.py} included with the arXiv post of this paper with $\texttt{N\_eq=10}$. Connected moments are denoted $x(i)$ and disconnected moments $d(i)$ (see Table \ref{tab:Graphs}).
{ \scriptsize 
\begin{align*}
 1 &=   x(1) + 3 g x(2) \\
0 &= 2  x(2) + 4 g   x(4) + 2 g  x(3) -2 x(1) -2  N^{-1}x(1)\\
0 &= 3  x(3) +  6g  x(7) +  3 g x(6) -3  N^{-1}d(1) -3 x(2) -3  N^{-1}x(2)\\
0 &= 3  x(4) + g   x(10) +  2 g x(8) +  4 g x(7) +  2 g x(9) -2  N^{-2}x(2) -2 x(2) -4  N^{-1}x(2) - d(1)\\
0 &= 3  x(5) + 9 g  x(11) -9 x(2)\\
0 &= 4  x(6) + 8 g   x(15) + 4 g   x(14) -8  N^{-1}d(3) -4 x(3) -4  N^{-1}x(3)\\
0 &= 4  x(7) + g   x(16) + 2 g  x(18) + 2 g  x(22) + 2 g   x(21) + 3 g   x(15) + g   x(19) + g  x(17)\\
&-2 x(4) -d(3) -3  N^{-1}d(3) -1 x(3) -3  N^{-1}x(4) -4  N^{-2}x(4) -2  N^{-1}x(3)\\
0 &= 4  x(8) + 2 g  x(28) + 4 g  x(16) + 2 g  x(27) +  2 g  x(21)\\
&+ 2 g x(23) -2 x(4) -2 d(3) -4  N^{-1}x(4) -6  N^{-2}x(4) -2  N^{-1}x(3)\\
0 &= 4  x(9) +2 g   x(22) + 2 g  x(27) + 4 g  x(17) +  2 g x(26)\\
&+ 2 g  x(23) -2 x(4) -4  N^{-2}x(3) -2 d(3) -6  N^{-1}x(4) -2  N^{-3}x(5)\\
0 &= 4  x(10) + 6 g  x(19) + 6 g  x(23) -3 x(4) -d(2) -6  N^{-2}x(3) -6  N^{-1}x(4).\\
\end{align*}
}

\begin{table}[H]
\begin{tabular}{ccccc}
\AppGr{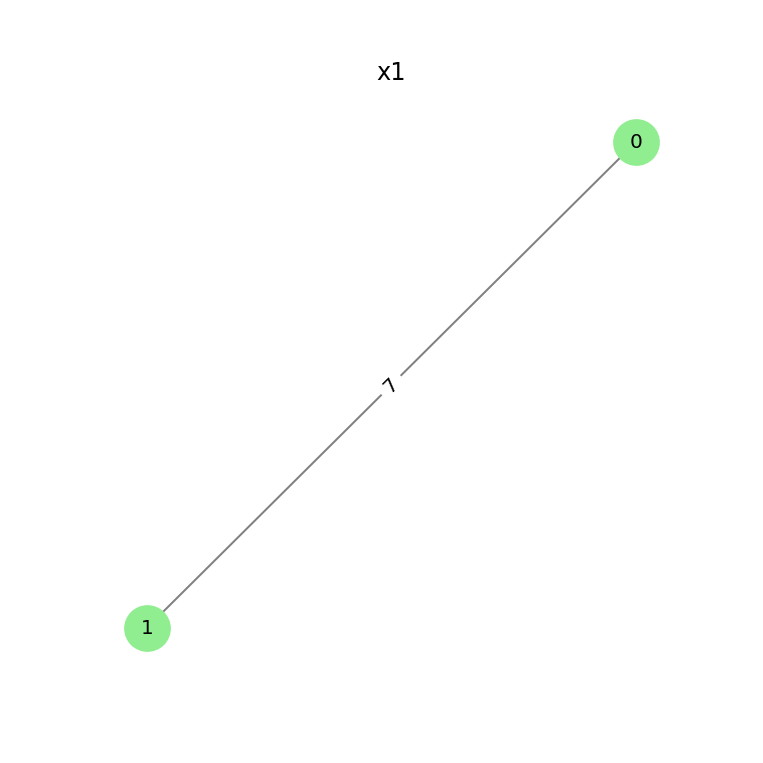} &
\AppGr{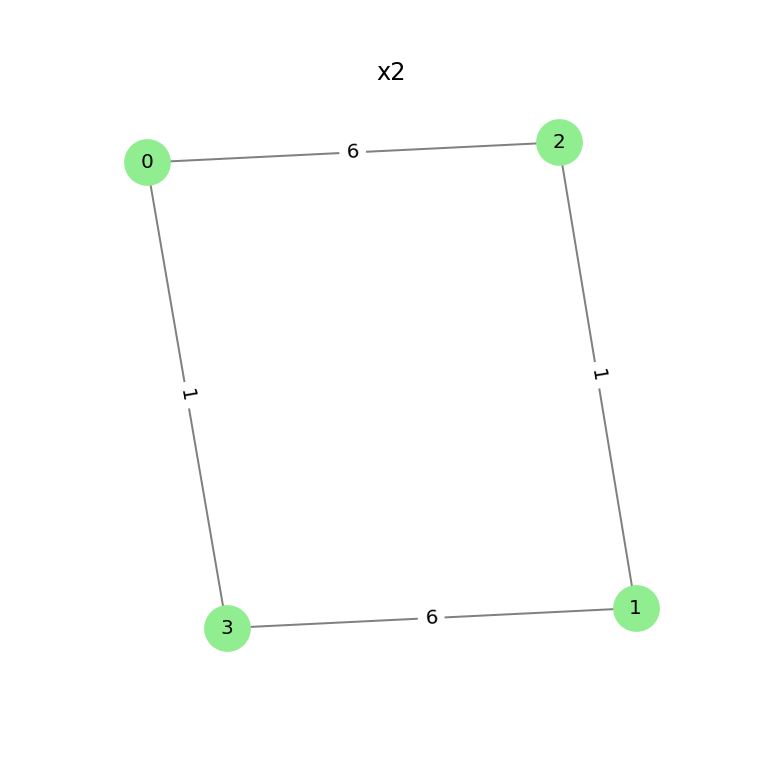} &
\AppGr{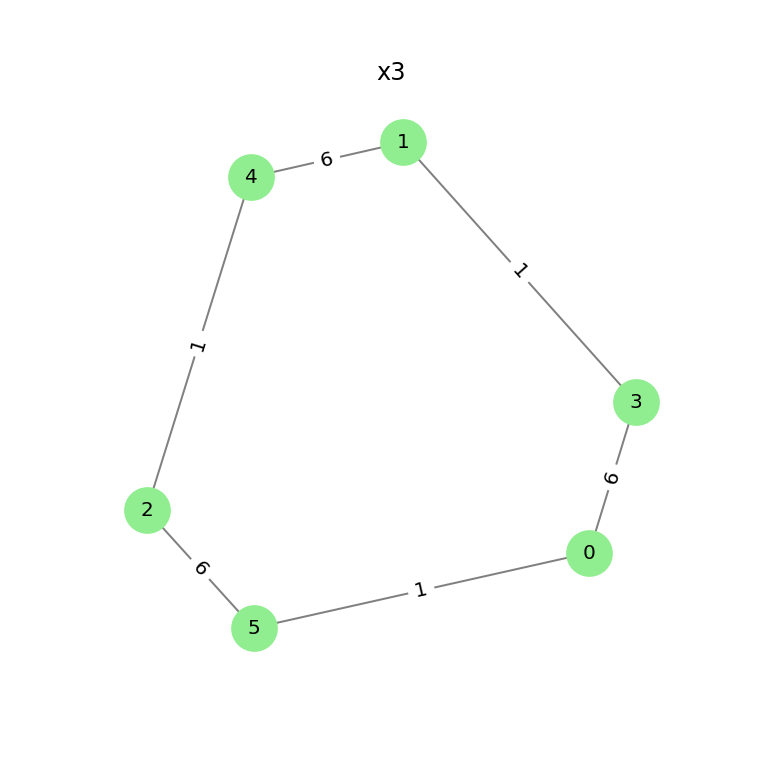} &
\AppGr{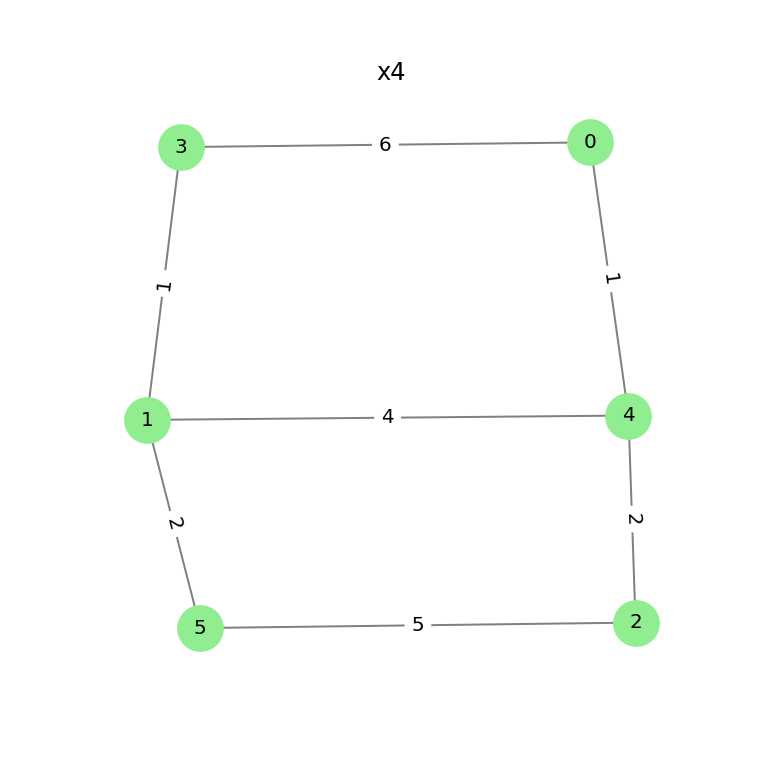} &
\AppGr{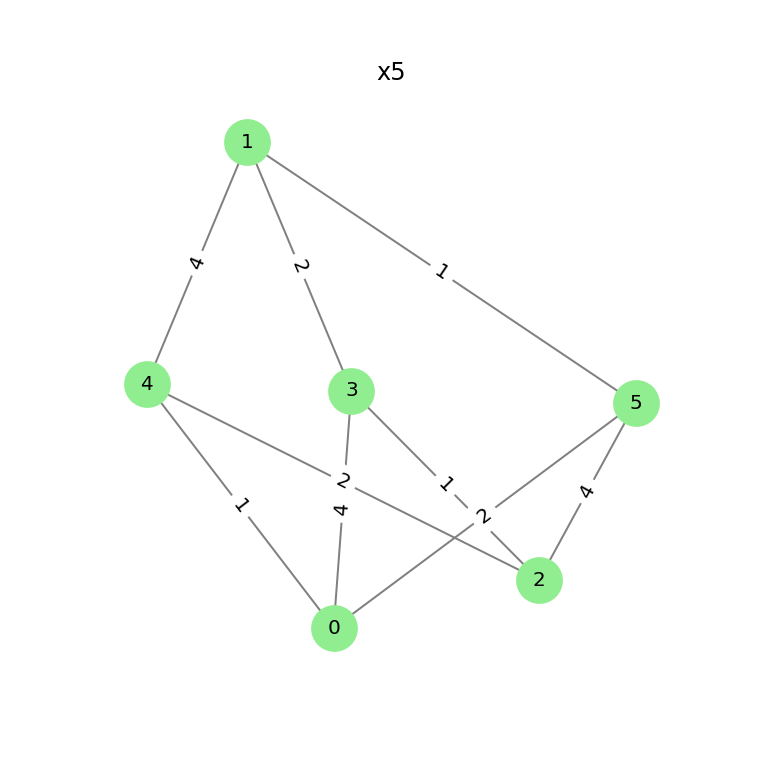} \\[-1ex]
\AppGr{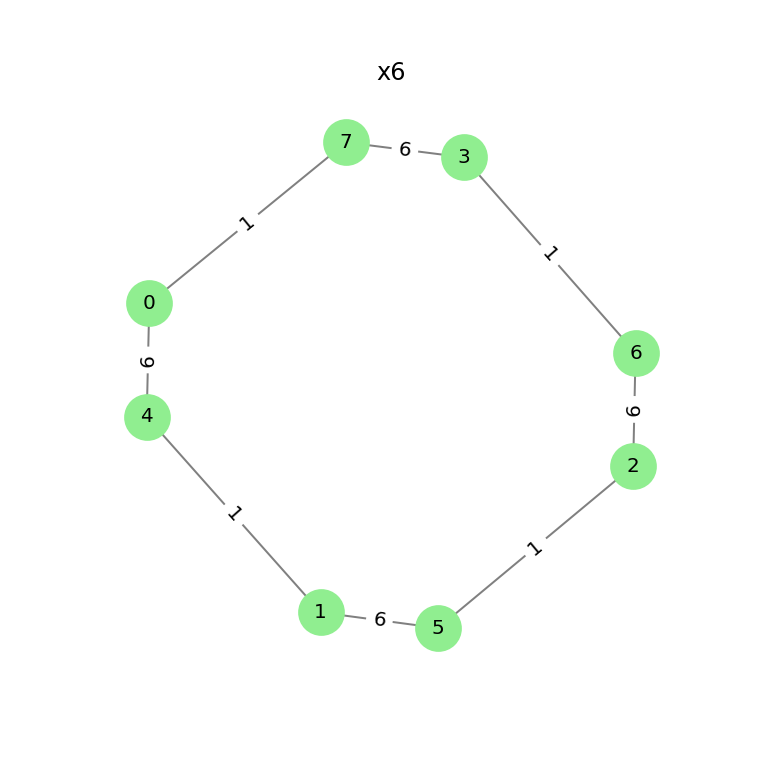} &
\AppGr{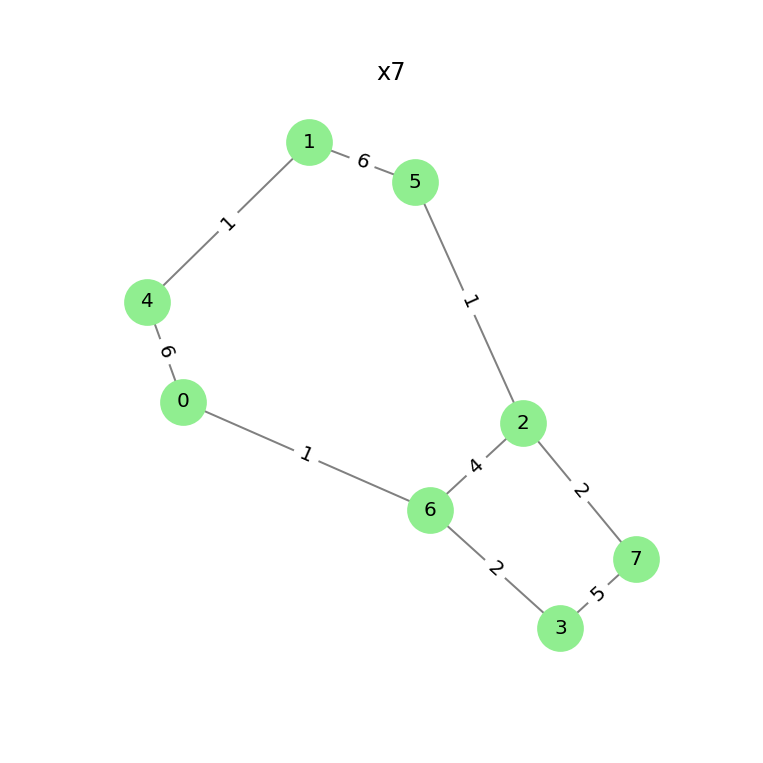} &
\AppGr{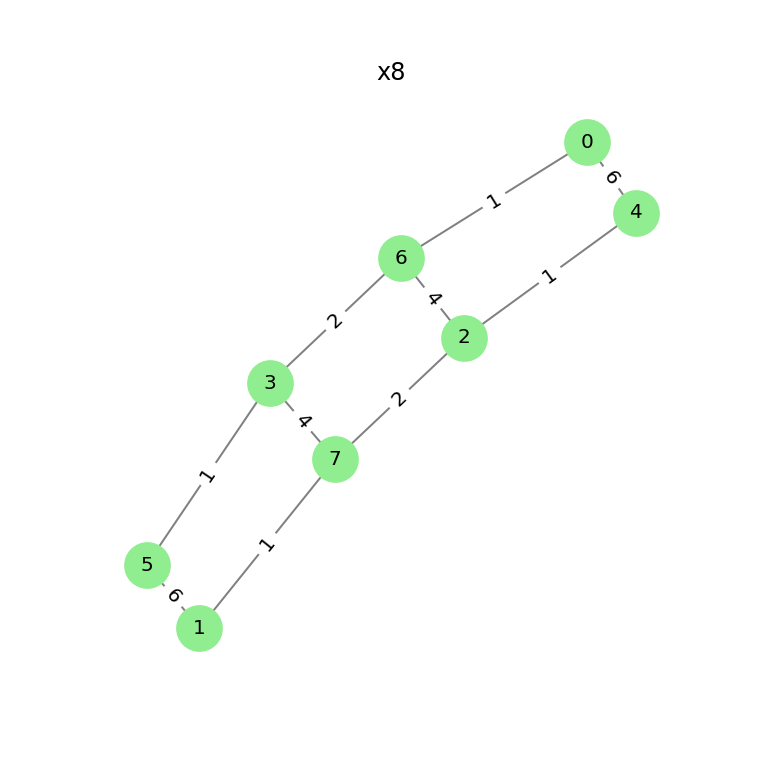} &
\AppGr{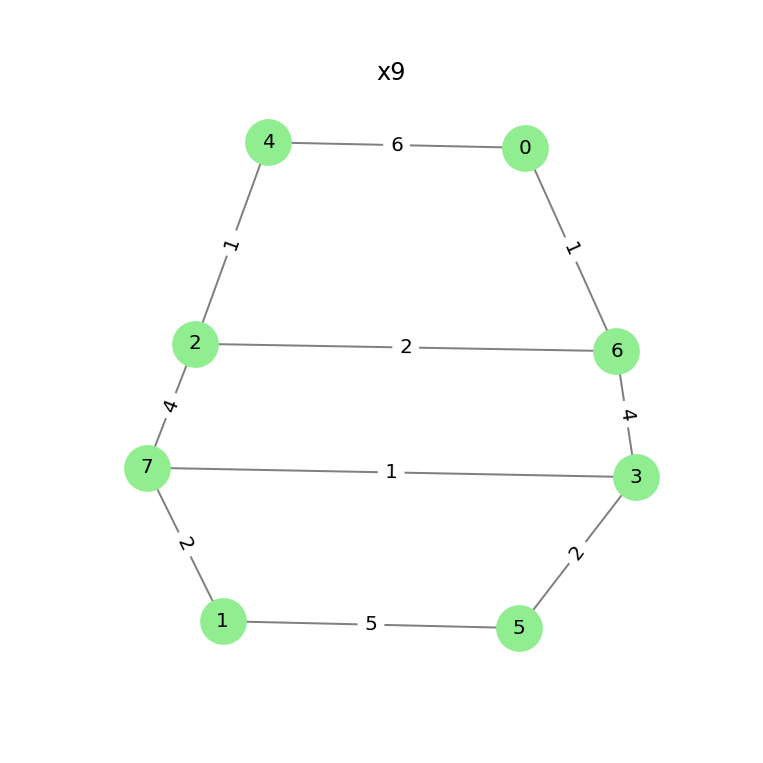} &
\AppGr{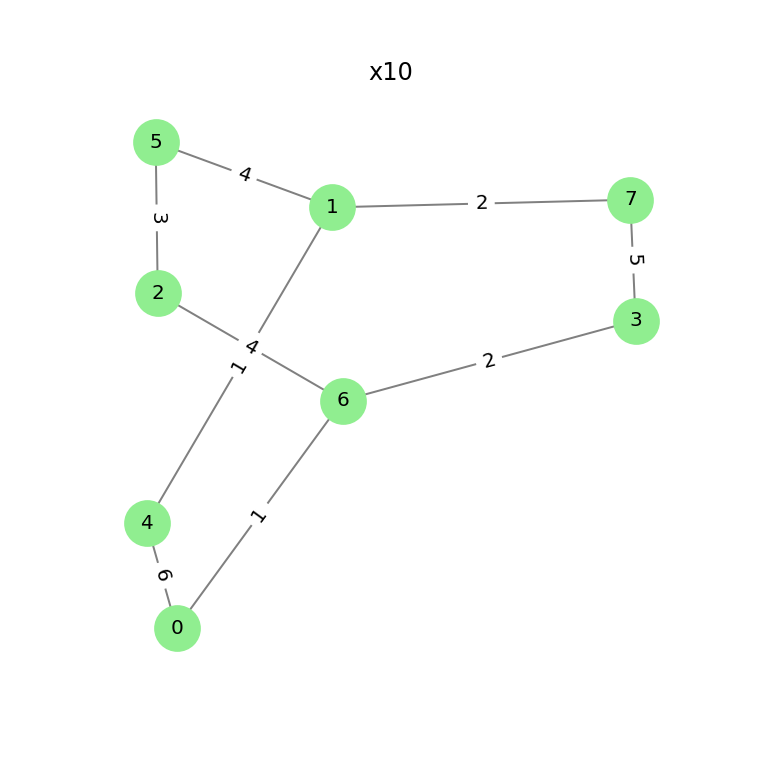}\\[-1ex]
\AppGr{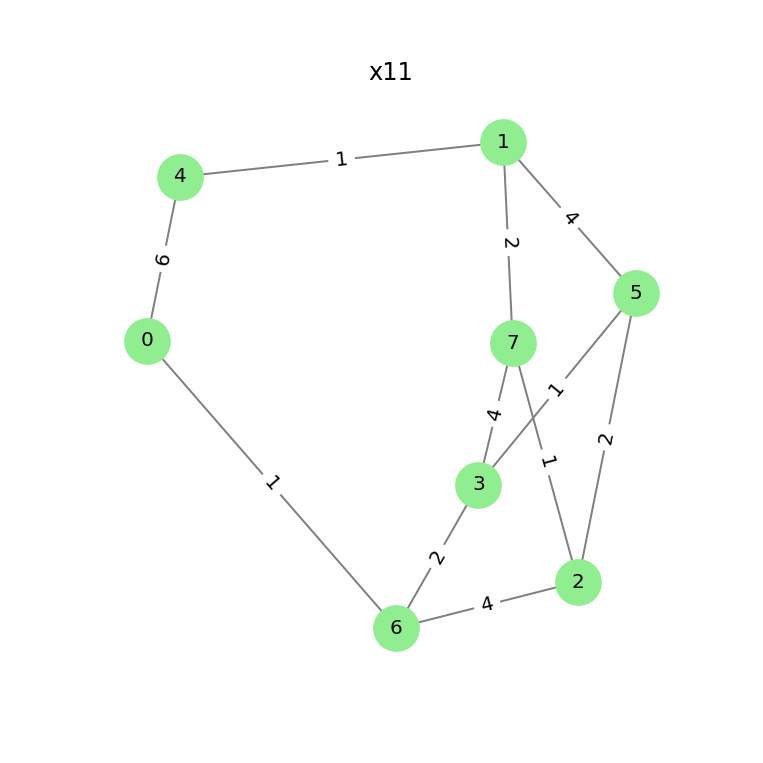} &
\AppGr{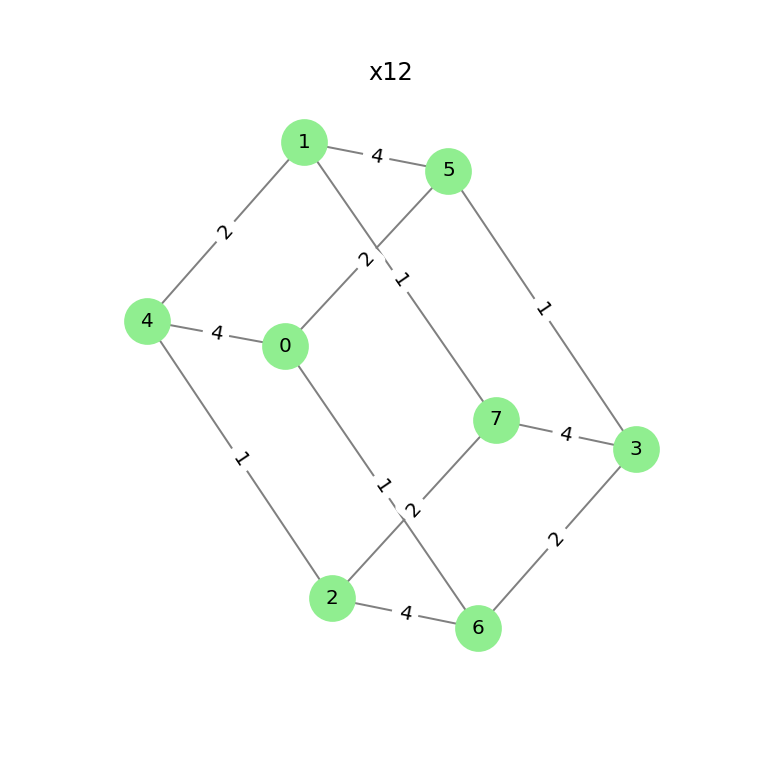} &
\AppGr{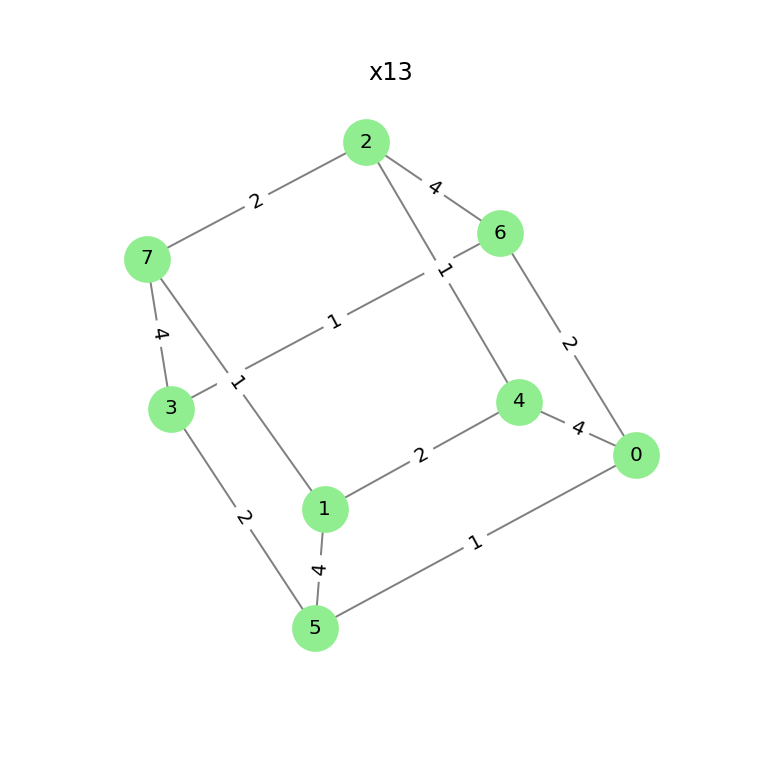} &
\AppGr{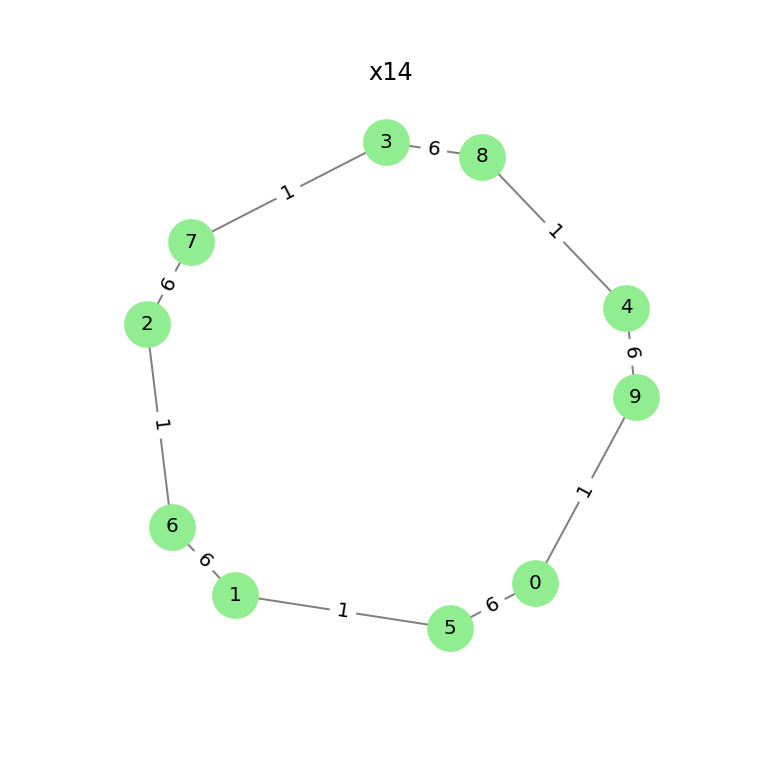} &
\AppGr{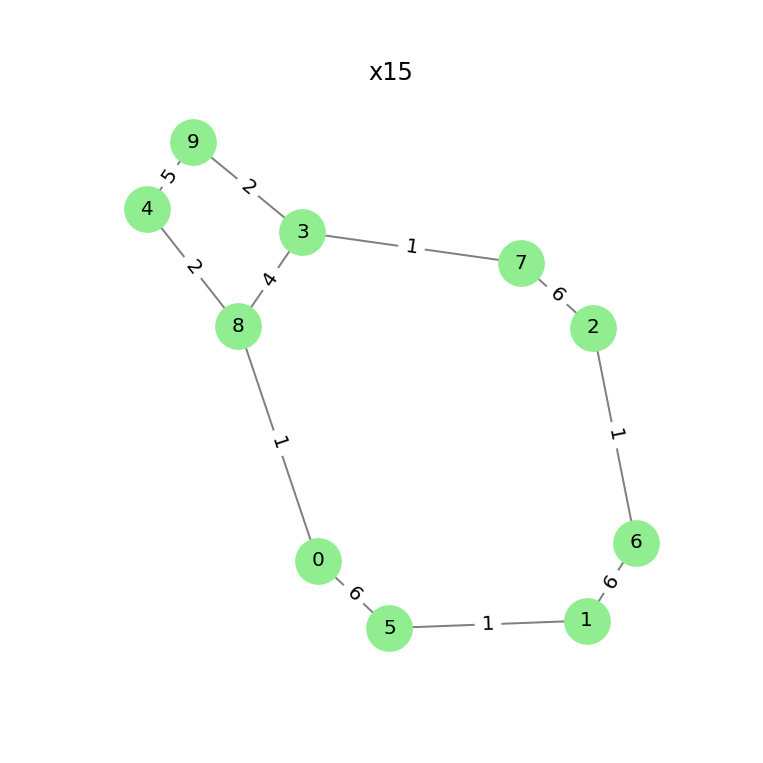} \\[-1ex]
\AppGr{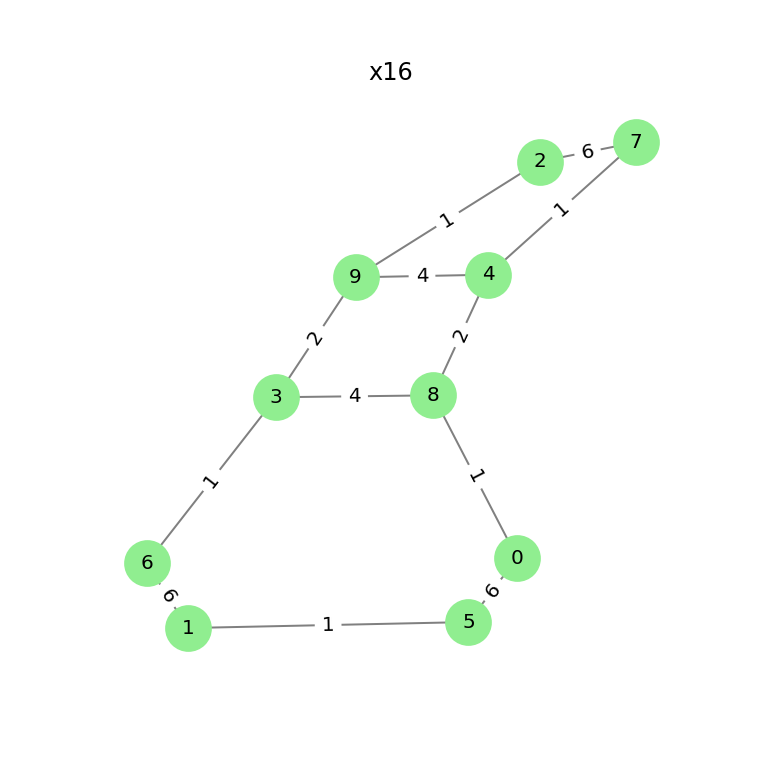} &
\AppGr{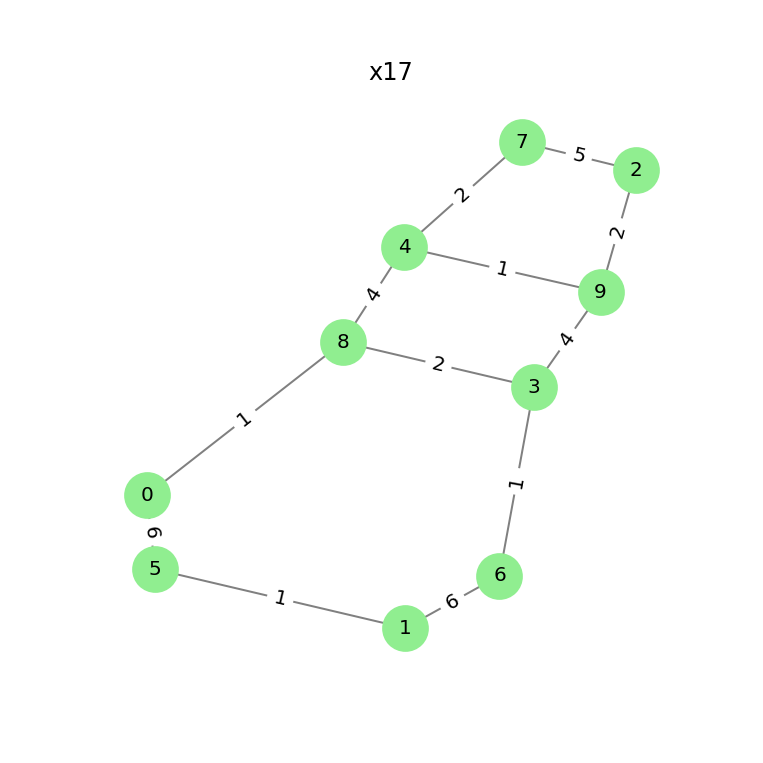} &
\AppGr{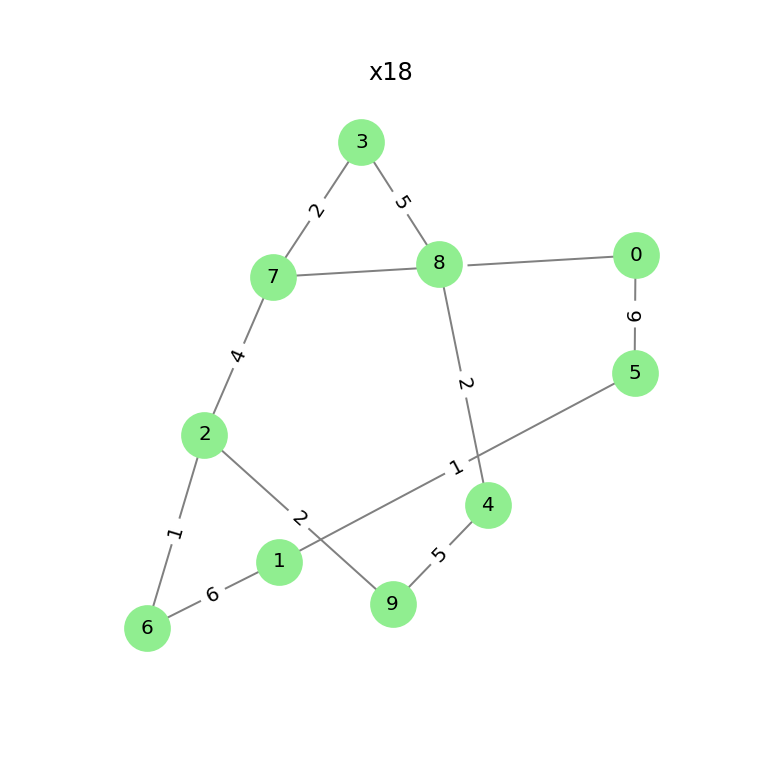} &
\AppGr{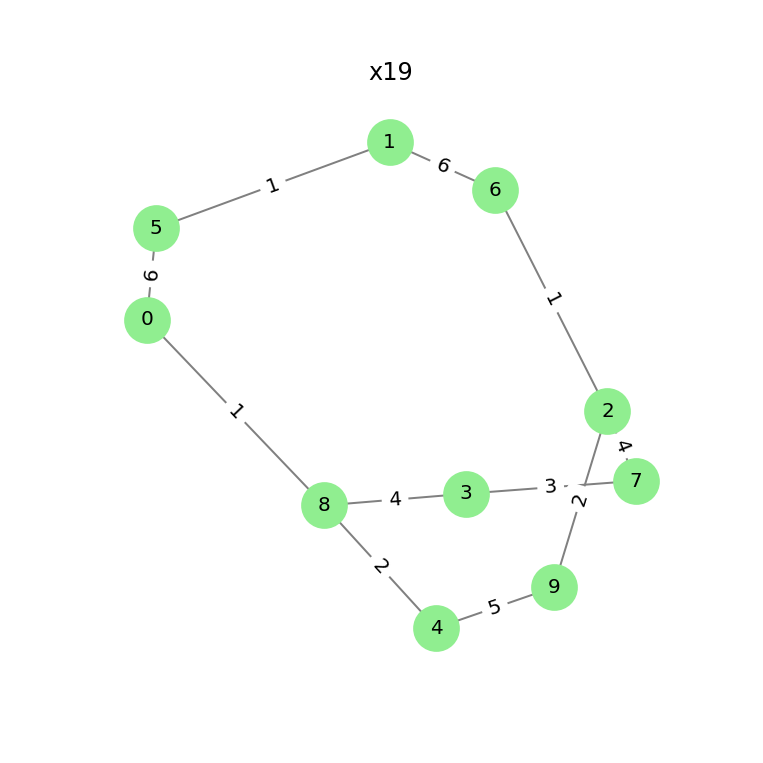} &
\AppGr{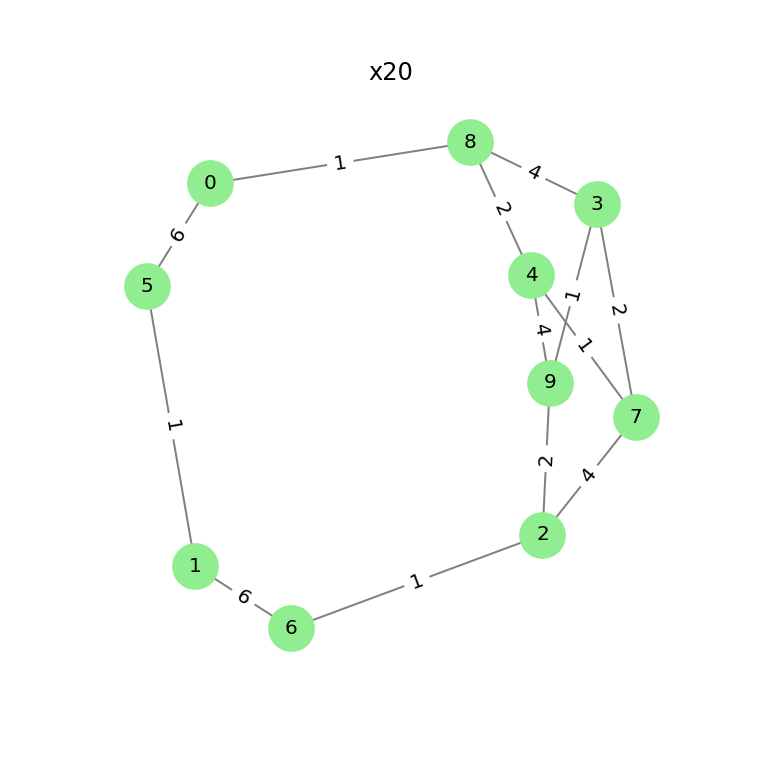} \\[-1ex]
\AppGr{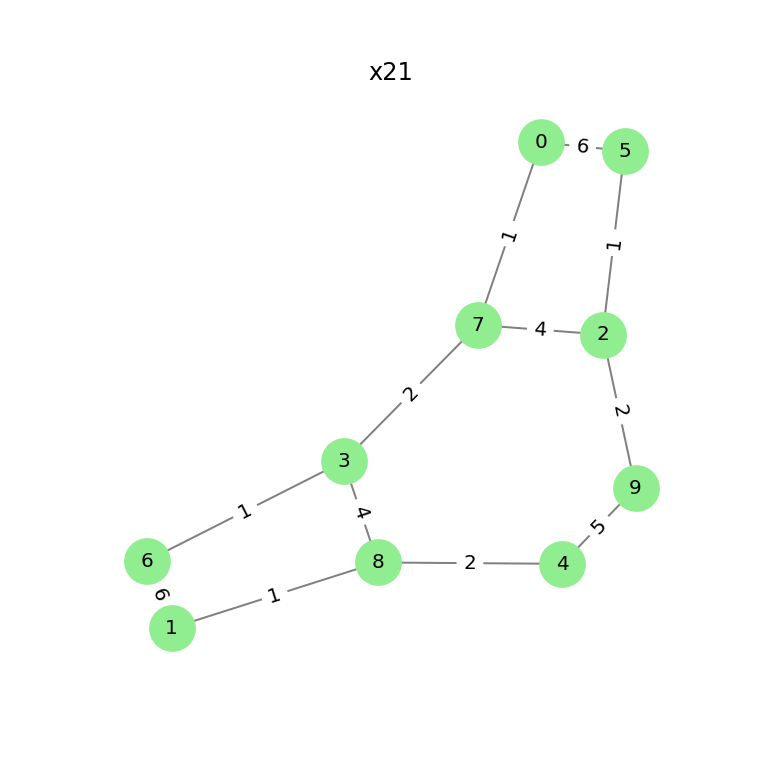} &
\AppGr{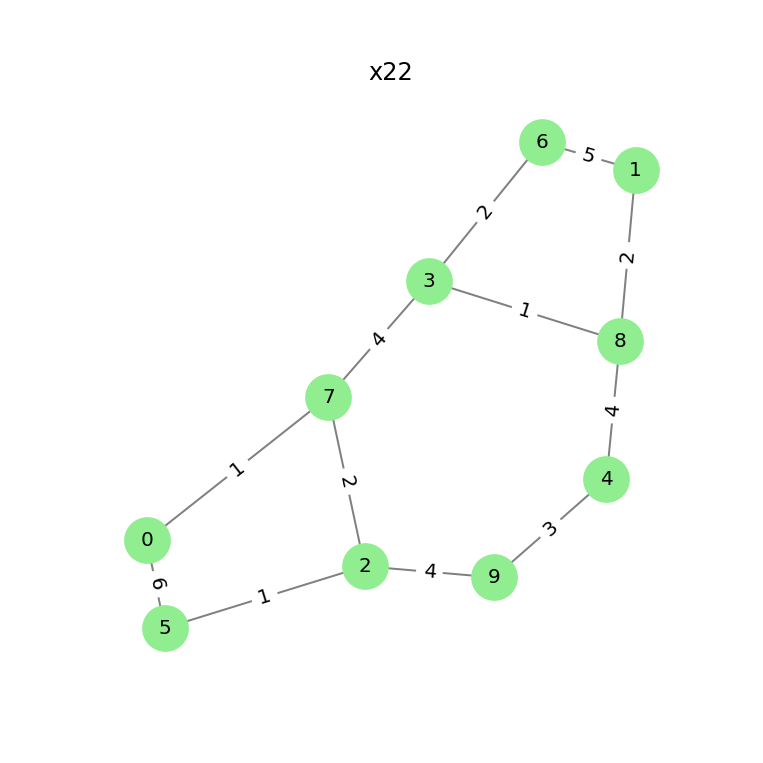} &
\AppGr{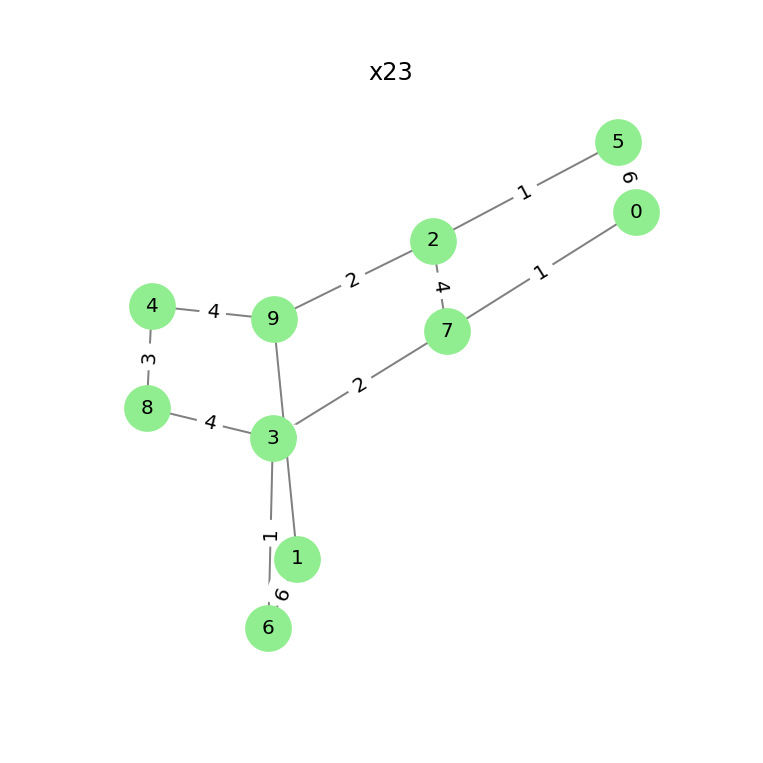} &
\AppGr{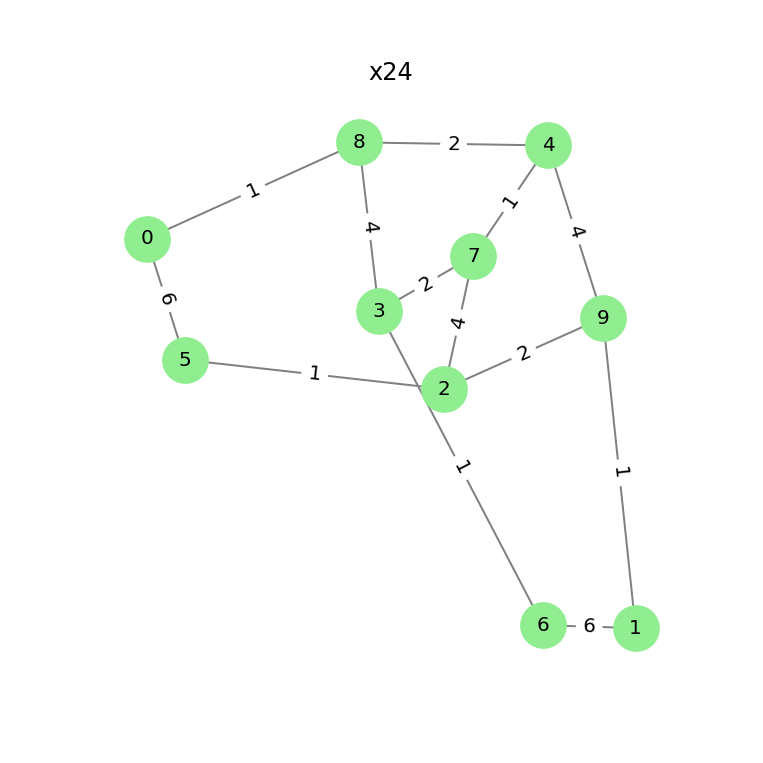} &
\AppGr{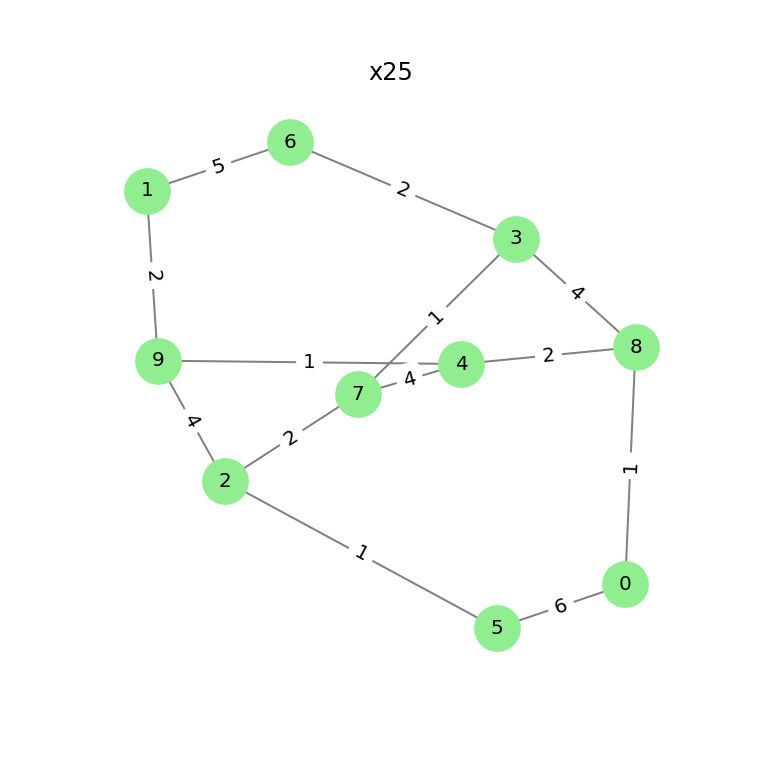} \\[-1ex]
\AppGr{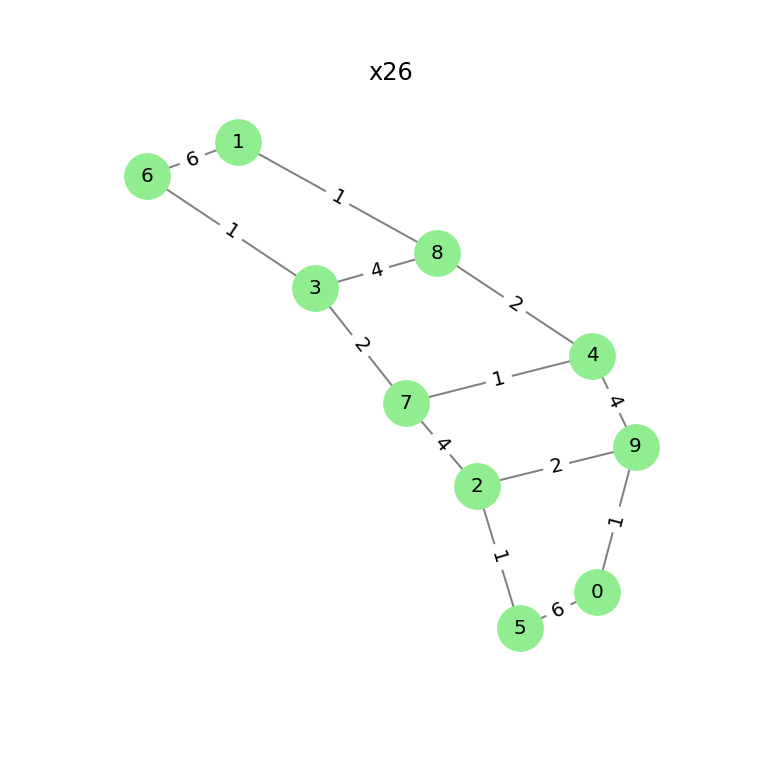} &
\AppGr{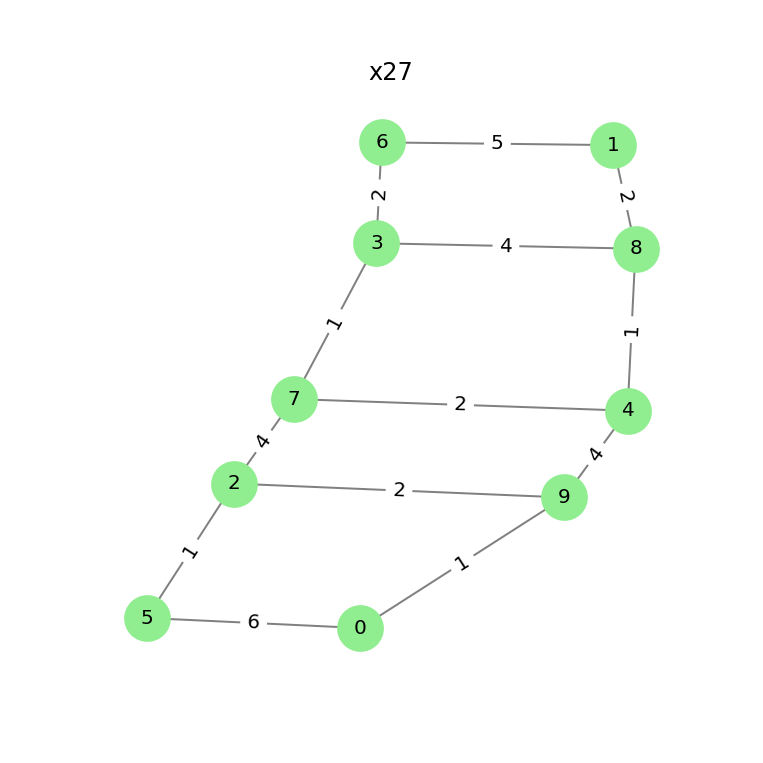} &
\AppGr{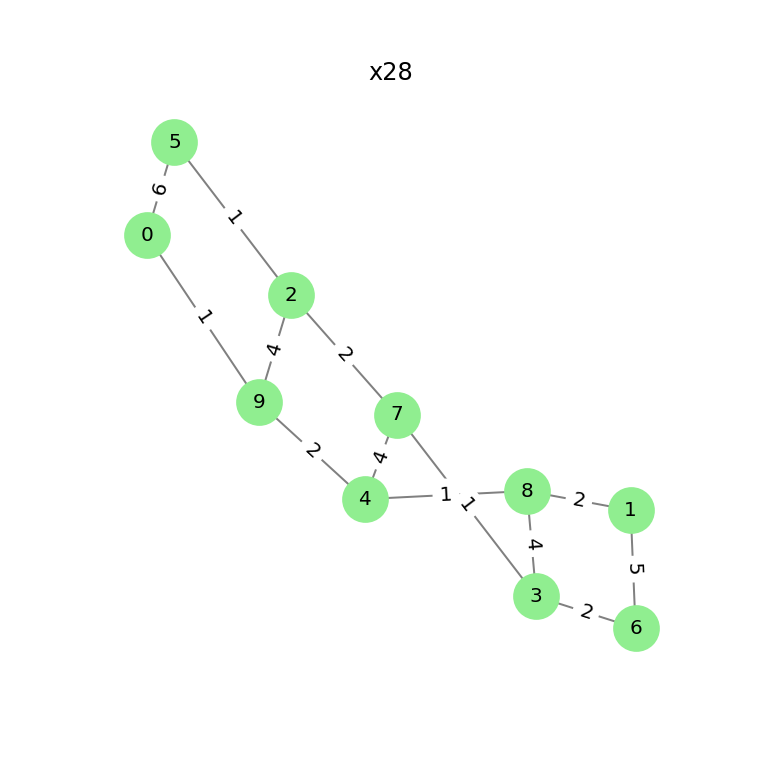} &
\AppGr{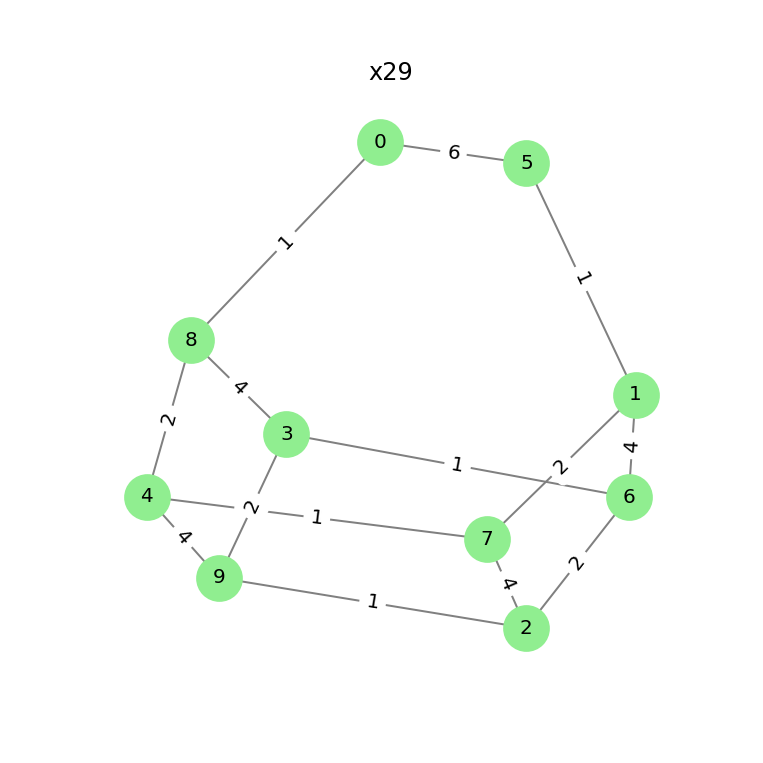} &
\AppGr{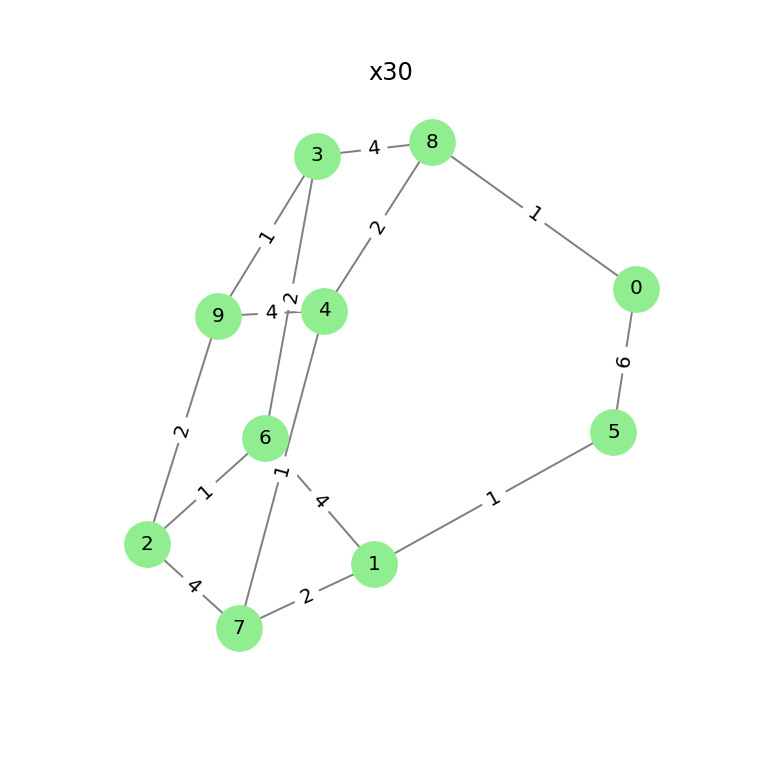} \\[-1ex]
\AppGr{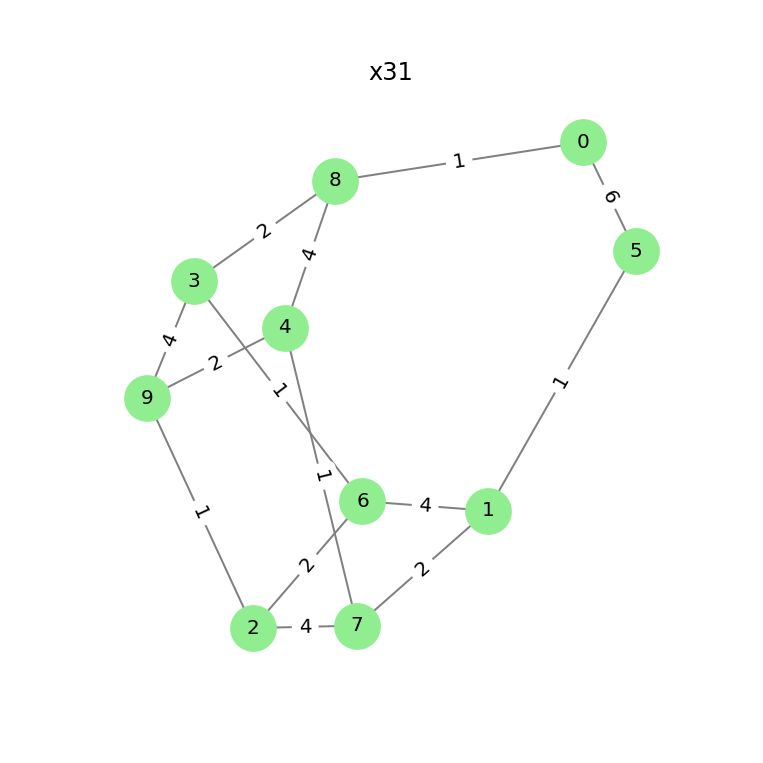} &
\AppGr{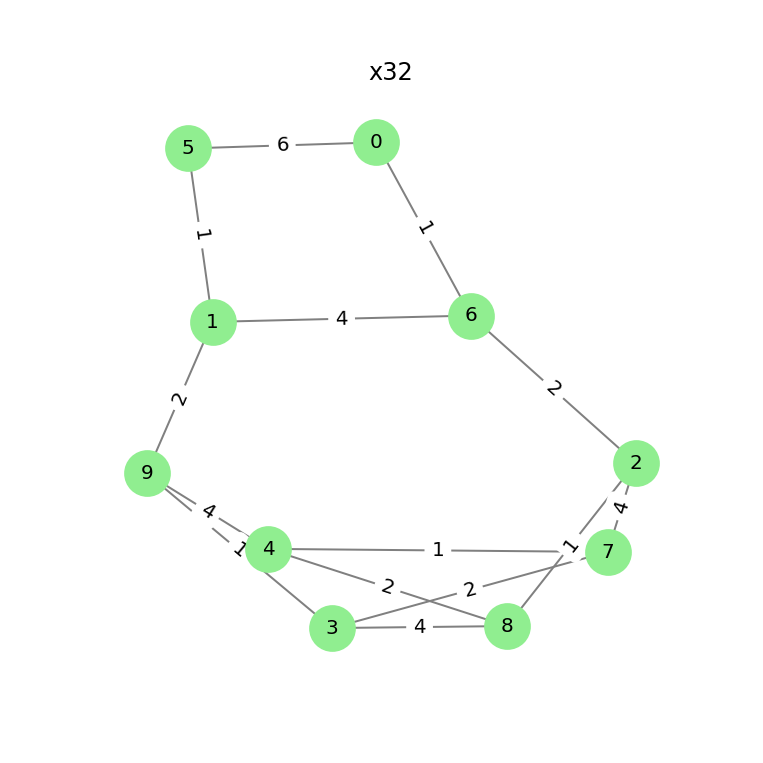} &
\AppGr{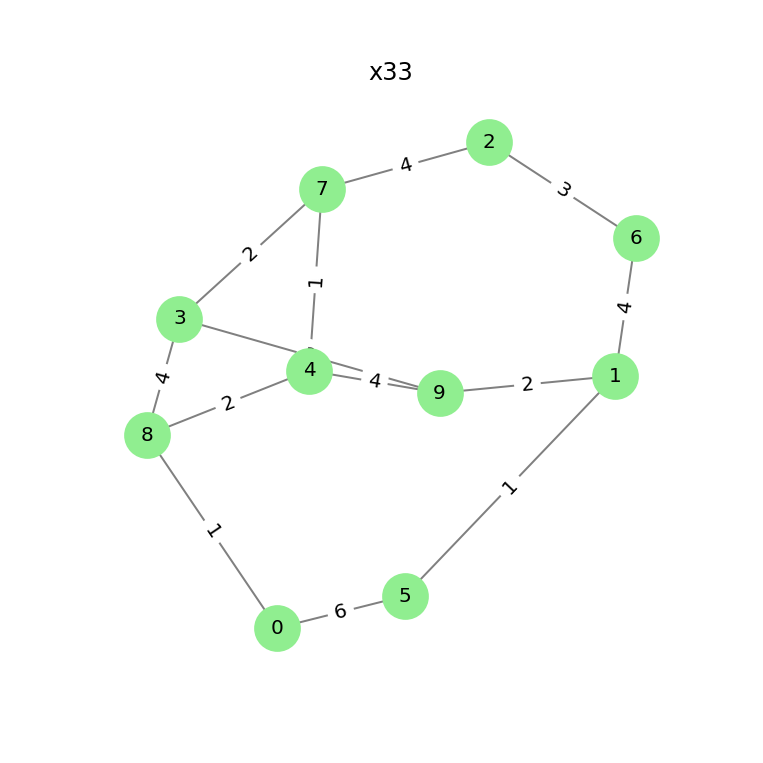} &
\includegraphics[height=2.8cm]{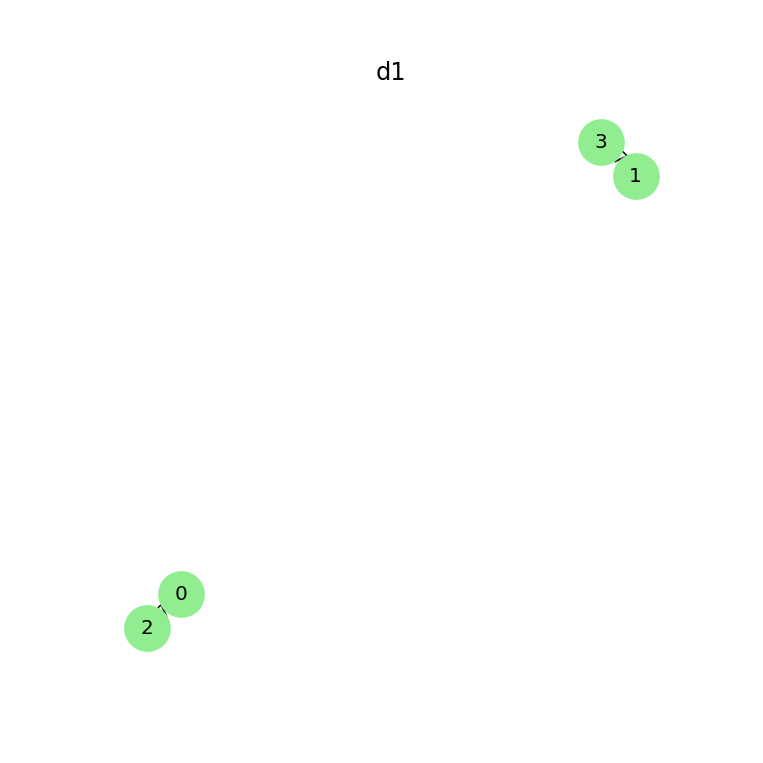}&
\includegraphics[height=2.8cm]{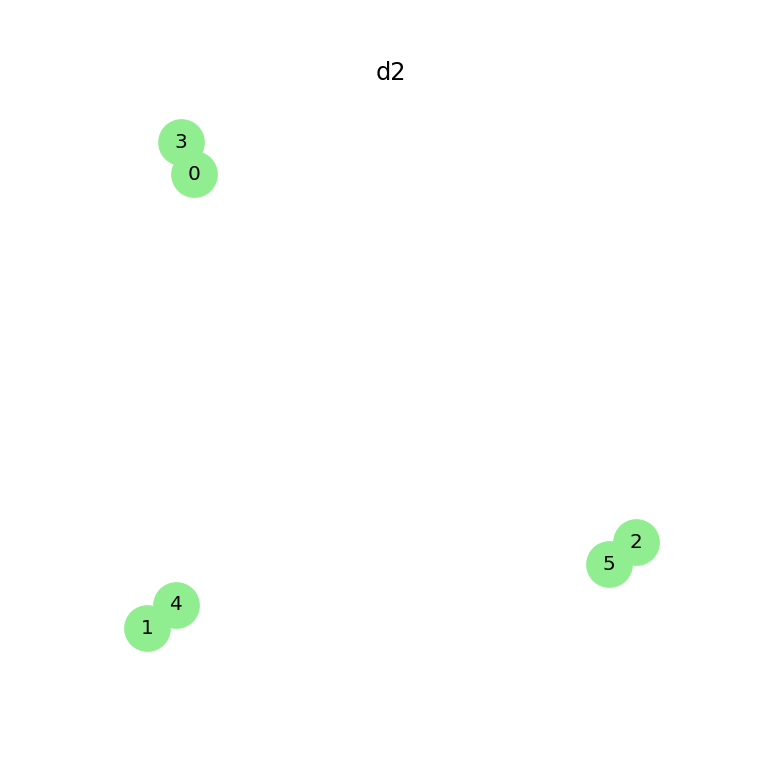}\\[-1ex] &
\includegraphics[height=2.8cm]{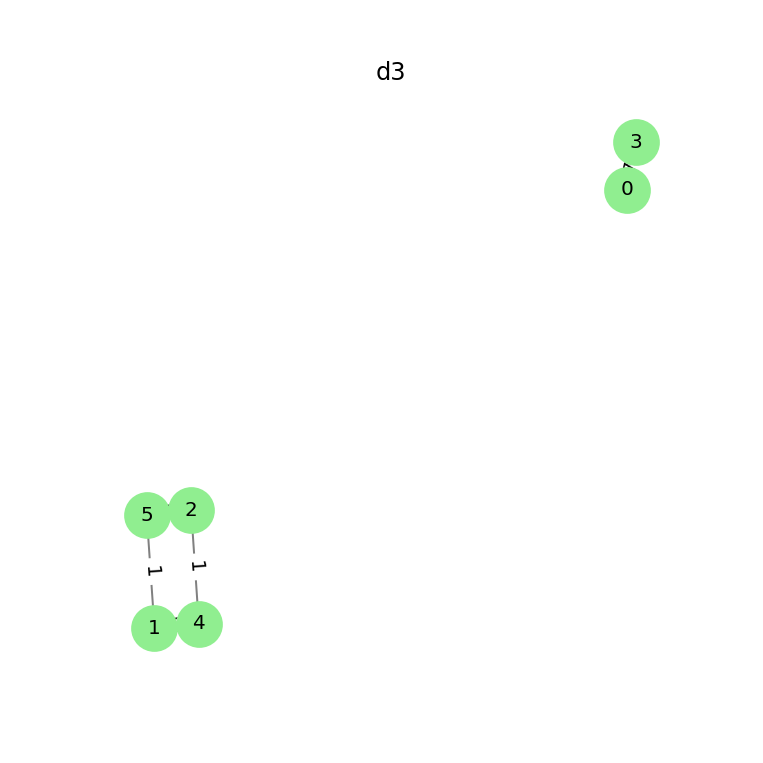}&&
\includegraphics[height=2.8cm]{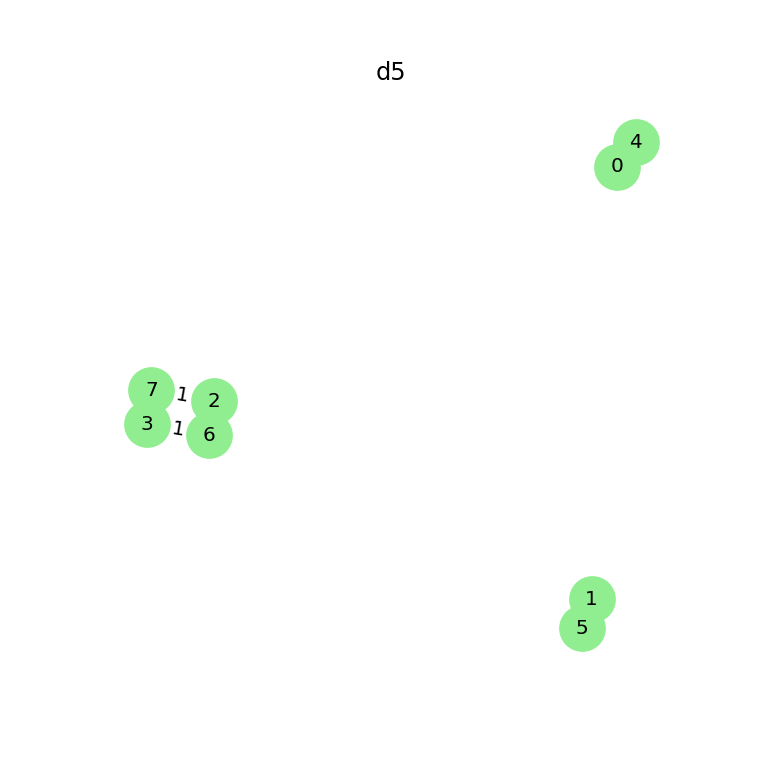} &
\end{tabular}

\caption{A table of graphs whose invariant appear in the first ten DSE of the quartic rank 3 model.
The  number edges marked on the visualization are determined by edge number and colouring
 with each colour here being denoted 1, 2, or 4. To avoid drawing double edges,
 we depict the colours' sum:  $6$ would denote edges
 of colour 2 and 4 connecting the same vertices, etc.
The $x(i)$  and $d(i)$ variables correspond to connected and disconnected graph $i$ respectively.  \label{tab:Graphs} }
\end{table}

\section{Double series of cumulants with a tensor integrator}\label{app:expansions}
\footnotesize
We give the double series for the two-point function for several models. The leading orders
stressed with bold fonts are the origin of the conjectures in Section \ref{sec:conjecture}.
These double series have been obtained with \texttt{feyntensor}.

\subsection{The quartic melonic model}
For the model
\[
S(T,\bar T)=
N^{2} \Big[
\littlemelon
+
\frac{g}{2}  \big(
\vone + \vtwo+ \vthree\big)
\Big] \,.
\]
explicit computation in \texttt{feyntensor}  yields
\[  \nonumber
 \mathbb E_g\conn [ \tfrac 1N \littlemelon ]
& = 1 - {3g}  \Big( 1+ \frac1N \Big)\\ \notag
&
+ {3g^2} \bigg[ \frac 1{2^2 \cdot 2!}  \Big(  \mathbf{16} + \frac{40}{N} + \frac{16}{N^{2}} + \frac{8}{N^{4}} \Big)+ \frac1{2^2}    \Big(\mathbf{16} + \frac{32}{N} + \frac{24}{N^{2}} + \frac{8}{N^{3}}   \Big) \bigg]
\\ \notag
& -  {g^3}\bigg[ \frac{3 }{2^3 \cdot 3! } \Big( \mathbf{240}+
\frac{1056}{N} + \frac{1056}{N^{2}} + \frac{240}{N^{3}} + \frac{480}{N^{4}} + \frac{480}{N^{5}}  \nonumber \\ \notag
&  +  \frac{6  }{2^3 \cdot 2!\cdot  1!  }  \Big( \mathbf{240}+ \frac{816}{N} + \frac{1056}{N^{2}} + \frac{720}{N^{3}} + \frac{480}{N^{4}} + \frac{240}{N^{5}}   \Big) \\ \notag
&  +  \Big( \mathbf{240}+ \frac{720}{N} + \frac{1008}{N^{2}} + \frac{1056}{N^{3}} + \frac{432}{N^{4}} + \frac{96}{N^{6}}
\Big) \bigg] \nonumber
\\ \notag\notag
& +  {g^4}\bigg[ \frac{3}{  2 ^ 4 \cdot  4! } \Big(\mathbf{5376} + \frac{35712}{N} + \frac{62976}{N^{2}} + \frac{35712}{N^{3}} + \frac{32256}{N^{4}} + \frac{64128}{N^{5}} + \frac{26880}{N^{6}} + \frac{8064}{N^{8}} \Big) \\ \notag  \nonumber &
+\frac{6}{2 ^ 4  \cdot 3! \cdot  1! }  \Big( \mathbf{5376} +\frac{27648}{N} + \frac{52608}{N^{2}} + \frac{54144}{N^{3}} + \frac{50688}{N^{4}} + \frac{45696}{N^{5}} + \frac{26880}{N^{6}} + \frac{8064}{N^{7}} \Big) \\ \notag   \nonumber&
+\frac{3}{2 ^ 4  \cdot 2! \cdot  2! }   \Big( \mathbf{5376}+ \frac{25088}{N} + \frac{50432}{N^{2}} + \frac{56960}{N^{3}} + \frac{53760}{N^{4}} + \frac{53504}{N^{5}} + \frac{21504}{N^{6}} + \frac{4480}{N^{8}}  \Big) \\ \notag  \nonumber &
+\frac{3}{2 ^ 4  \cdot 2! \cdot  2! }    \Big( \mathbf{5376} +
\frac{23296}{N} + \frac{46336}{N^{2}} + \frac{64512}{N^{3}} + \frac{63232}{N^{4}} + \frac{38016}{N^{5}} + \frac{20608}{N^{6}} + \frac{9728}{N^{7}}
\Big) \bigg]  \\ \notag \notag
& -  \Big(\frac{g}{2}\Big)^5\bigg[
\frac{3}{5!} \Big( \mathbf{161280}+ \frac{1482240}{N} + \frac{3955200}{N^{2}} + \frac{3955200}{N^{3}} + \frac{3095040}{N^{4}} + \frac{6766080}{N^{5}} \\ \notag &  + \frac{6604800}{N^{6}} + \frac{1612800}{N^{7}} + \frac{1854720}{N^{8}} + \frac{1854720}{N^{9}}  \Big) +
\frac{6}{1!\cdot 4!} \Big( \mathbf{161280}+
\frac{1159680}{N} + \frac{3045120}{N^{2}}
\\ \notag\notag
& + \frac{4277760}{N^{3}} + \frac{4915200}{N^{4}} + \frac{5959680}{N^{5}} + \frac{5694720}{N^{6}} + \frac{3548160}{N^{7}} + \frac{1854720}{N^{8}} + \frac{725760}{N^{9}}
\Big)
\\ \notag\notag
& +
\frac6{2! \cdot 3! } \Big(\mathbf{161280}+
\frac{1008000}{N} + \frac{2720640}{N^{2}} + \frac{4258560}{N^{3}} + \frac{5253120}{N^{4}} + \frac{6533760}{N^{5}} \\ \notag & + \frac{5992320}{N^{6}} + \frac{3064320}{N^{7}} + \frac{1543680}{N^{8}} + \frac{806400}{N^{9}}  \Big)
+\notag
\frac3{1! \cdot 1! \cdot 3! } \Big(  \mathbf{161280}+
\frac{961920}{N} + \frac{2520960}{N^{2}} \\ \notag&
+ \frac{4348800}{N^{3}} + \frac{5965440}{N^{4}} + \frac{6088320}{N^{5}} + \frac{5040000}{N^{6}} + \frac{4272000}{N^{7}} + \frac{1706880}{N^{8}} + \frac{276480}{N^{10}}
\Big)\notag \\ \notag\notag
& +
\frac{3}{1!\cdot 2! \cdot 2!}
 \Big( \mathbf{161280}+
 \frac{898560}{N} + \frac{2378240}{N^{2}} + \frac{4350720}{N^{3}} + \frac{6067200}{N^{4}} + \frac{6458880}{N^{5}} + \frac{5391360}{N^{6}} \\ \notag&  + \frac{3375360}{N^{7}} + \frac{1672960}{N^{8}} + \frac{587520}{N^{9}}
\Big)
\bigg] \notag
+ O(g^6)\, \\
& =1 - 3 g+ 18 g^{2} - 135 g^{3} + 1134 g^{4} - 10206 g^{5}  + \ldots \notag \\ \notag&  + (-3g  + 39 g^{2} - 462 g^{3} + 5367 g^{4} - 62058 g^{5}+\ldots ) \frac 1N \notag \\ \notag& + (24 g^{2} - 588 g^{3} + 10488 g^{4} - 164520 g^{5}+\ldots ) \frac1 {N^2}  \notag \\ \notag& + (6 g^{2} + 417 g^{3} + 12381 g^{4} + 272970 g^{5}+\ldots ) \frac1 {N^3}  \notag \\ \notag& + (3 g^{2} - 264 g^{3} + 11868 g^{4} - 358308 g^{5}+\ldots ) \frac1 {N^4}  \notag \\ \notag& + (-120 g^{3} + 9429 g^{4} -  400446 g^{5}+\ldots ) \frac1 {N^5}   \notag \\ \notag& + (- 12 g^{3} + 4830 g^{4} - 348390 g^{5}+\ldots ) \frac1 {N^6}  \notag \\ \notag&  + (1416 g^{4} - 222720 g^{5}+\ldots ) \frac1 {N^7}    + (273 g^{4} - 105939 g^{5}+\ldots ) \frac1 {N^8} \notag \\ \notag&  -( 33489 g^{5}+ \ldots) \frac1 {N^9} - (4320 g^{5}+ \ldots) \frac1 {N^{10}}+ O(N^{-11}, g^6)
\]

\subsection{Model with sextic cyclic melonic interactions}
The next sextic model is also known as `cyclic melonic',
\[
S_q(T,\bar T)=
N^{2} \bigg[
\littlemelon
+  \frac q3 \big( \Qone + \Qtwo + \Qthree \big)  \bigg ]   \,.
\]
Using the Tensor Integrator  \texttt{feyntensor}
\[ \nonumber
\mathbb E_q\conn [ \tfrac 1{ N} \littlemelon ] & =1 -
\frac{q}{3}
\Big [ 3 \Big (3+ \frac{9}{N} + \frac{3}{N^{2}} + \frac{3}{N^{4}} \Big )
\Big]
 \notag \\ \notag
 & + \notag
\Big(\frac{q}{3}
\Big)^2
\bigg[ \frac{3}{2!}\Big(\mathbf{54}+
\frac{432}{N} + \frac{828}{N^{2}} + \frac{432}{N^{3}} + \frac{504}{N^{4}} + \frac{1188}{N^{5}} + \frac{450}{N^{6}} + \frac{216}{N^{8}}  \Big)\\ \notag &+  \frac{3}{2!}
\Big( \mathbf{54}+
\frac{324}{N} + \frac{702}{N^{2}} + \frac{648}{N^{3}} + \frac{720}{N^{4}} + \frac{1080}{N^{5}} + \frac{396}{N^{6}} + \frac{180}{N^{8}}
\Big) \notag
\bigg]
\\ &-
\Big(\frac q3 \Big)^3
\bigg[
\frac{1}{3!} \Big(\mathbf{
1944}+
\frac{17496}{N} + \frac{67068}{N^{2}} + \frac{149688}{N^{3}} + \frac{249318}{N^{4}} + \frac{411156}{N^{5}} \\ \notag & \notag+ \frac{583848}{N^{6}}  + \frac{533628}{N^{7}} + \frac{449550}{N^{8}} + \frac{464616}{N^{9}} + \frac{176904}{N^{10}} + \frac{47952}{N^{12}} \Big)\\ \notag & \notag
+\frac{3}{3!}
\Big(\mathbf{1944}+
\frac{28188}{N} + \frac{118584}{N^{2}} + \frac{188568}{N^{3}} + \frac{173016}{N^{4}} + \frac{408240}{N^{5}} \\ \notag & \notag+ \frac{690768}{N^{6}}  + \frac{380052}{N^{7}} + \frac{285768}{N^{8}} + \frac{571536}{N^{9}} + \frac{231336}{N^{10}} + \frac{75168}{N^{12}}
\Big)\\ \notag & \notag
+\frac{6}{3!}
\Big(  \mathbf{1944}+
\frac{20412}{N} + \frac{81162}{N^{2}} + \frac{161352}{N^{3}} + \frac{228420}{N^{4}} + \frac{401436}{N^{5}}  \\ \notag & + \frac{607986}{N^{6}} + \frac{506412}{N^{7}} + \frac{414072}{N^{8}} + \frac{486972}{N^{9}} + \frac{188568}{N^{10}} + \frac{54432}{N^{12}}
\Big) \bigg]
\]

\subsection{Model with sextic non-cyclic melonic interactions}

\[
S_\gamma(T,\bar T)=
N^{2} \Big[
\littlemelon
+  \gamma  \big( \Eone+ \Etwo + \Ethree \big)  \Big ]   \,.
\]
Explicit computation using the Tensor Integrator  \texttt{feyntensor} yields:
\[ \nonumber
\mathbb E_e\conn [ \tfrac 1{ N} \littlemelon ] & =1 -
\gamma  \bigg[ 3 \cdot\Big
(3 +\frac{6}{N} + \frac{6}{N^{2}} + \frac{3}{N^{3}}\Big )
\bigg]
\\ \notag
& +
\gamma^2 \bigg[
\frac{3}{2!}
\Big(\mathbf{54}+
\frac{240}{N} + \frac{558}{N^{2}} + \frac{816}{N^{3}} + \frac{912}{N^{4}} + \frac{996}{N^{5}} + \frac{444}{N^{6}} + \frac{84}{N^{8}}  \Big)
\\ \notag
 &+\frac{3}{2!}  \Big(\mathbf{54}+ \frac{228}{N} + \frac{516}{N^{2}} + \frac{888}{N^{3}} + \frac{1056}{N^{4}} + \frac{744}{N^{5}} + \frac{426}{N^{6}} + \frac{192}{N^{7}}
\Big)
\bigg]
\\ \notag & -
\gamma^3 \bigg[
\frac{1}{3!}
\Big( \mathbf{ 1944}+   \frac{12636}{N} + \frac{44388}{N^{2}} + \frac{119718}{N^{3}} + \frac{253584}{N^{4}} + \frac{422010}{N^{5}} \notag
  \\  &  +    \frac{585252}{N^{6}} + \frac{621270}{N^{7}} + \frac{499608}{N^{8}} + \frac{400950}{N^{9}} + \frac{168912}{N^{10}} + \frac{22896}{N^{12}}
\Big)
\\ \notag
 &+\frac{3}{3!}  \Big(\mathbf{1944} +
 \frac{13752}{N} + \frac{51066}{N^{2}} + \frac{125676}{N^{3}} + \frac{239022}{N^{4}} + \frac{410508}{N^{5}} \notag
  \\  \nonumber& + \frac{587466}{N^{6}} + \frac{632520}{N^{7}} + \frac{534618}{N^{8}} + \frac{338184}{N^{9}} + \frac{162468}{N^{10}} + \frac{55944}{N^{11}}
\Big)\\ \notag
 &+\frac{6}{3!}  \Big(\mathbf{1944}+
 \frac{13032}{N} + \frac{46548}{N^{2}} + \frac{121536}{N^{3}} + \frac{251640}{N^{4}} + \frac{419796}{N^{5}} \notag
  \\  &  +  \frac{566568}{N^{6}} + \frac{632268}{N^{7}} + \frac{556632}{N^{8}} + \frac{330984}{N^{9}} + \frac{153252}{N^{10}} + \frac{58968}{N^{11}}
\Big)
\bigg]\nonumber
\]

\normalsize

\section{Implementation of graph operations for the DSE}\label{app:implementation}

Here we take from \cite{gehrig} to
encode a $D$-coloured graph $G$ in the computer, instead of using the
adjacency matrix. That reference automated in Python, to some
extent, the functional renormalization group approach to tensor
models.  These renormalization tools are not used, but some of the graph operations
turned out to be useful here to construct the DSEs. 

\begin{remark}
The thesis \cite{gehrig} was \textit{not} used in \texttt{feyntensor} \cite{feyntensor}, which instead is implemented on the SageMath \texttt{Graph} class.
\end{remark}
Replacing the adjacency matrix, we define another one, $A_G \in M_{D
  \times p} (\{1,\ldots, p\}) $,
where $p$ is the number of black or white vertices, $p=\#
V(G)/2$. Choosing any numeration of such sets, and fixing any colour
$c$, the colour-$c$ edge pairing in $G$, $ \sigma_c: \Vb(G)\to
\Vw(G)$, will be understood as a permutation $\sigma_c\in \Sym(p)$ (the
$k$-th black vertex is matched with the $\sigma_c(k)$-th white vertex
by the colour-$c$ edge).  Let
\[
A_G=[ \sigma_{c}(v)  ]_{ \substack{ c =1,\ldots, D ; v\in \Vb(G) } }
\]
($A_G$ depends on the numerations $\alpha: \Vw(G)\to \{1,\ldots,p\} $ and $\gamma
: \Vb(G)\to \{1,\ldots,p\}$, under which
this matrix will be transformed to
$A_G'= [ \alpha\circ \sigma_{c}\circ \gamma\inv (v)  ]_{ \substack{ c =1,\ldots, D ; v\in \Vb(G) } } $).
To match Python's  convention, the colour index $c$ runs through the set $\{0,\ldots, D-1\}$
instead of  $\{1,\ldots, D\}$ as above. Hence these matrices read in the code as follows,
\[
\runter{\includegraphics[width=8cm]{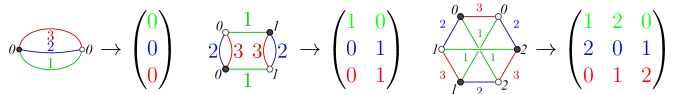}}
\]
(The previous picture is taken from \cite{gehrig}; the italic fonts enumerate vertices).

 \newcommand{\etalchar}[1]{$^{#1}$}
\bibliographystyle{alpha}

\end{document}